\documentclass[]{spie}  

\usepackage{amsmath,amsfonts,amssymb}
\usepackage{multirow}
\usepackage{graphicx}
\usepackage{setspace}
\usepackage{tocloft}
\usepackage[normalem]{ulem} 
\usepackage[colorlinks=true, allcolors=blue]{hyperref}

\graphicspath{{./}{Figures/}}

\title{High-contrast aperture masking interferometry: detecting companions within the diffraction limit}

\author[a]{Benjamin Calvin} 
\author[a]{Michael Fitzgerald} 
\author[a]{Jessica Castellanos} 
\affil[a]{University of California - Los Angeles, Physics and Astronomy, 430 Portola Plaza, Los Angeles, CA, USA, 90095}

\authorinfo{Further author information: (Send correspondence to B.C.)\\B.C.: E-mail: bcalvin@astro.ucla.edu, ben.a.calv@gmail.com}

\begin{document} 
\maketitle

\begin{abstract}
The pursuit of the direct discovery and characterization of Earth-like exoplanets has motivated high contrast imaging surveys using coronagraphs with narrow inner working angles.  Traditional imaging methods, however, are limited by the size of the on-axis PSF and typically only detect companions at angular separations of $>2 \lambda/D$. Meanwhile, non-redundant masking (NRM) techniques have been used with single-aperture telescopes to interferometrically resolve astrophysical features at angular scales within the diffraction limit, but NRM interferometry faces sensitivity issues for high-contrast objects. In this project, we investigate utilizing high contrast coronagraphy in conjunction with NRM interferometry. 
We conduct a preliminary investigation of this combined architecture with the goal of determining the potential bounds of performance and discovery space for high-contrast targets within the inner working angle of traditional high-contrast imaging. We predict an improvement in the achievable contrast for photon-limited observations by combining the techniques, and we also develop estimates for wavefront stability requirements for the system.
\end{abstract}


\section{Introduction} \label{sec:intro}  
The detection and characterization of an Earth-like exoplanet in the habitable zone (HZ) of a Sun-like star is a leading priority assigned from the Astro2020 Decadal survey\cite{2021pdaa.book.....N}. To accomplish this, the measurements and analysis of
an observation must overcome over ten orders of magnitude of flux contrast between the on-axis starlight and the reflected light from the companion. The angular separation between the nearest 
stars and their respective habitable zones is approximately 50\,milli-arcseconds (mas),\cite{2021Msngr.182...38K} which serves as an angular scale at which to achieve the high-
contrast goal.

One of the technologies being developed for the purpose of directly studying exoplanets at close separations is coronagraphy; vortex coronagraphy, in particular, has been investigated for achieving high contrast at a small inner working angle (IWA) \cite{2005ApJ...633.1191M, 2005OptL...30.3308F, 2008MNRAS.384..515J, 2009OExpr..17.1902M}. By utilizing an optical phase ramp of a given optical charge in an up-stream focal plane, light from an on-axis source can be removed while allowing for the transmission of off-axis light into the scientific instrument. 
The IWA of the coronagraph increases with a higher-charged vortex, restricting the region of close-separation that can be imaged, but the high charge makes the system more resilient against tip-tilt phase residuals which degrade the contrast achievable by low-charge vortex coronagraphs.


Traditional high-contrast imaging techniques typically constrain the discovery space 
of detectable off-axis companions to well outside the coronagraph's inner working angle 
($> 2 \lambda/D$). Between differential imaging causing PSF self-subtraction
\cite{2014ApJ...780...25E}, limited adaptive optics performance\cite{2019A&A...631A.106S, 2022A&A...665A.136P}, and classical angular resolution constraints, most high-contrast technologies are designed to survey for off-axis targets 
at radial separations beyond multiple angular resolution elements from their host. 
To image a companion with 50\,mas separation from its host star at $3\,\lambda/D$, the optical system would require a 10\,m diameter telescope to operate at a wavelength of $\leq 1$\,micron (or, equivalently $\leq 500$\,nm on a 5\,m telescope or in the \textit{L} band for an ELT).
Modern vortex coronagraphs can achieve a raw contrast of $10^{-4}$ on-sky \cite{2024AJ....168...78W}, 
while, in optical testbeds, the combination of coronagraphs and advanced wavefront control algorithms (e.g. electric field conjugation\cite{2007SPIE.6691E..0AG}, linear dark field control \cite{2017JATIS...3d9002M}, self-coherent cameras\cite{2006dies.conf..553B, 2020A&A...635A.192P}, etc.) have been able to dig dark holes that reach contrast levels of $10^{-8}$ at separations of $\geq 3\lambda /D$.\cite{2017JATIS...3d9002M, 2019SPIE11117E..1FR, 2020A&A...635A.192P, 2022SPIE12180E..25B, 2026arXiv260322551D}. While the vortex coronagraph can suppress photon flux from on-axis sources to improve the off-axis contrast in the final focal plane, the scale of the inner-bound of the discovery space ultimately constrains the science cases that can be investigated with these technologies. 

Although traditional imaging techniques face limitations at close-separation regions of interest, other methods (e.g. aperture masking) are capable of distinguishing astrophysical sources separated at even smaller field angles. 
Aperture masking interferometry (AMI) is a technique that transforms a single-aperture telescope into an interferometric array, where, instead of the science detector measuring an image, it measures a combination of overlapping fringes (an "interferogram"). 
When the mask is a non-redundant mask (NRM), meaning that the baseline between any two subapertures of the mask is distinct from every other baseline, the resulting interferometer is capable of measuring astrophysical features from each baseline, uniquely\cite{1958MNRAS.118..276J, 1986Natur.320..595B, 2000PASP..112..555T, 2011A&A...532A..72L}. 
Thus, the amplitude, phase, and visibility of each baselines' resulting fringe will encode spatial information from the source onto distinct locations in the Fourier transform of the interferogram, which can then be measured from the "interferometric observables" (e.g. squared visibility, closure phase, biphase, etc.)\cite{1958MNRAS.118..276J, 1999PhDT........19M}. Because of the non-redundancy in the baselines of the aperture mask, these observables can be used to detect features on spatial scales smaller than the classical diffraction limit of the clear-aperture telescope. 
NRM interferometers have been used to detect stellar and sub-stellar companions, as well as extended disks, at sub-$\lambda/D$ separations \cite{2006ApJ...650L.131L, 2008ApJ...678L..59I, 2010A&ARv..18..317A, 2011A&A...532A..72L, 2019AJ....157..249G}. Detections from NRM surveys have been restricted to moderate contrast objects ($\Delta \text{mag} \approx 7$) as the interferometry signal scales with the source contrast, making high-SNR measurements of high-contrast objects challenging\cite{2011A&A...532A..72L, 2019JATIS...5a8001S}.

The strengths of AMI include the ability to detect astrophysical signals at the smallest achievable separations, but AMI can be limited in the detectable contrast at these separations. The vortex coronagraph has the ability to suppress on-axis light while allowing transmission for off-axis sources at close separations. However, traditional imaging techniques prevent the sources at those small separations from being distinguished from the on-axis source. 
The scope of this paper is to initially investigate the combination of vortex coronagraphy and NRM interferometry into a vortex+NRM (vNRM) system inside a numerical simulation with the goal of combining the strengths of both techniques. Ideally, the vortex coronagraph will occlude the bright on-axis light from a star, improving the off-axis contrast to better than would be detectable by the NRM alone. Meanwhile, the NRM architecture will allow for the detection of the high-contrast companions at closer separations than would be allowed by the coronagraph alone. Overall, by combining the two systems, this may create a new method for observing high-contrast targets at $\leq \lambda/D$ separations. 

In Sec. \ref{sec:broad_sim_methods}, we introduce the formalism we use to describe the vNRM. This includes the description of our conceptual-level simulation and the observables that will be measured by it. 
In Sec. \ref{sec:Photon_noise}, we introduce photon noise and simulate contrast curves in photon-limited measurements. 
In Sec. \ref{sec:WFE_general}, we introduce static and dynamic wavefront error to determine the robustness of the interferometric observables against them. 
Then, we discuss the implications of our findings in different observing environments and outline opportunities for future development 
in Sec. \ref{sec:discussion}, before summarizing our key takeaways in Sec. \ref{sec:conclusion}.

\begin{figure}[t] 
\centering
\includegraphics[width = 0.95\textwidth]{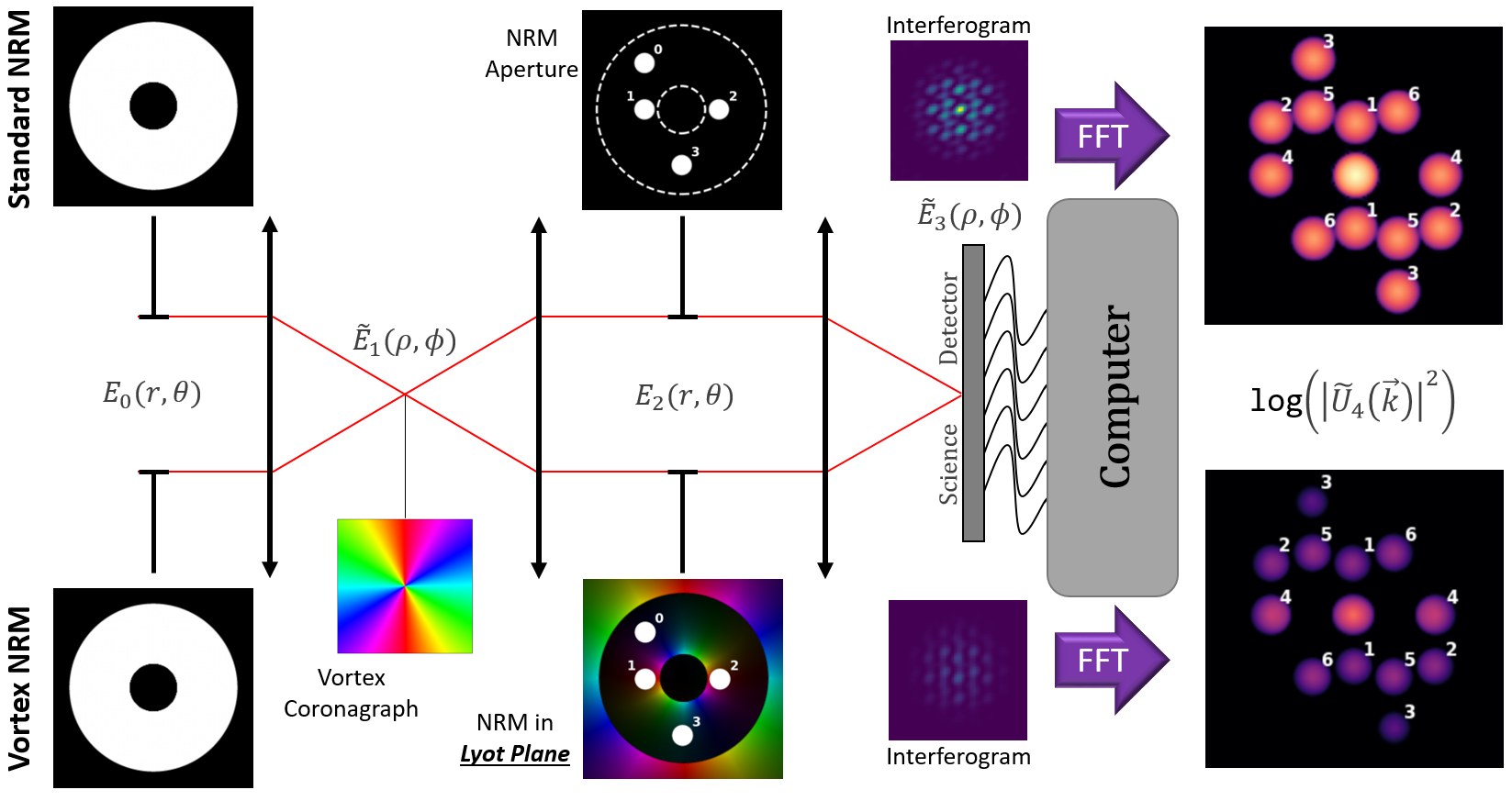}
\caption{Block diagram of the standard NRM (top) and charge-2 vNRM architecture (bottom) with annotated subaperture indices. In $\widetilde{E}_3$, the vNRM interferogram for an on-axis source is fainter than in NRM. Likewise, in $\widetilde{U}_4$, the squared magnitude of the UV plane shows less power each splodge (annotated by baseline index) for an on-axis wavefront.} \label{fig:block_diagram} 
\end{figure}

\section{Conceptual Design and Simulation Setup} \label{sec:broad_sim_methods}
A vortex coronagraph uses a vortex phase mask in a focal plane in conjunction with a pupil mask in the following Lyot plane to suppress on-axis light while allowing off-axis light to be transmitted. A NRM separates an individual pupil into an interferometric array that can allow for the detection of astrophysical features within traditional imaging limits. 
In this paper, we numerically simulate a vortex coronagraph combined with a NRM imaging system. 
We show a block diagram describing the NRM and vNRM systems in Fig. \ref{fig:block_diagram}, which begins with an initial telescope aperture shared between the two systems. In the standard NRM case (top), the NRM is applied in a pupil plane conjugate to the entrance plane and the resulting wavefront is then brought to a focus to image the resulting interferogram with a detector. In the vNRM case (bottom), we 
place the NRM in the Lyot plane of a vortex coronagraph and then focus the subsequent wavefront onto a detector. 

In this section, we provide a mathematical background to describe the fundamental elements of the combined vNRM system. 
Then, we describe the properties of the simulation that defines the vNRM system at the conceptual level. 
We, then, describe the analysis techniques to measure astrophysical signals from the NRM and vNRM systems, and finally, we compare the NRM and vNRM observables in the case of a point-source star with a point-source off-axis companion. 

\subsection{Analytical Framework} \label{sec:math}
In describing the vNRM system, we adopt the formalism used in Sauvage et al. (2010)\cite{2010JOSAA..27A.157S} to describe the significant wavefront planes. 
The optical system comprises a telescope aperture, coronagraphic focal plane, the proceeding pupil plane (the Lyot plane), and a detector plane. 
The initial telescope plane is indexed with $0$. The focal plane, where a scalar vortex coronagraph of even charge is applied, is indexed by $1$. The Lyot plane, in which the NRM is applied, is indexed by $2$. This is then brought to the final focal plane, indexed by $3$. In this plane, the interferogram is recorded by the detector.

The complex amplitude of the electric fields in the pupil planes is denoted by $E_0(r, \theta)$ and $E_2(r, \theta)$, with $r$ and $\theta$ being the radial and azimuthal spatial variables, respectively. The complex amplitudes in the focal planes are denoted by $\widetilde{E}_1(\rho, \phi)$ and $\widetilde{E}_3(\rho, \phi)$, with $\rho$ and $\phi$ being the radial and azimuthal variables describing the location in the focal plane. The "pre" and "post" subscripts refer to the planes prior and subsequent to the application of a mask in the optical system. 
In this work, we neglect all dependencies of the complex amplitudes to any polarization effects. 

We use Fraunhofer diffraction to describe the propagation of the complex amplitudes between successive planes such that $\widetilde{E}_1(\rho, \phi) = \text{FT}^{-1}\{E_0(r, \theta)\}$\cite{2010JOSAA..27A.157S}, where $\text{FT}^{-1}$ is the inverse Fourier transform. 
The general complex amplitude at the entrance aperture of the telescope is:
\begin{equation}
    E_0(r, \theta) = \mathcal{P}(r, \theta) \times \mathcal{S}(r, \theta)\text{ exp}[i(\psi_\text{dyn} + \psi_\text{stat})],
    \label{eqn:E0_entrance}
\end{equation}
where $\mathcal{P}$ is the telescope pupil function and $\mathcal{S}$ describes the amplitude errors and scintillation in the entrance plane. 
$\psi_\text{dyn}$ and $\psi_\text{stat}$ are dynamic and static phase aberrations in the telescope pupil plane, respectively.  %

The complex amplitude in the first focal plane is then:
\begin{equation}
    \widetilde{E}_{1, \text{pre}}(\rho, \phi) = \text{FT}^{-1}\{E_0(r, \theta)\} = \iint E_0 \text{ exp}[-i\rho r\text{cos}(\phi-\theta)]rdrd\theta 
    \label{eqn:E1_preVC}
\end{equation}
Then, at this focus, the vortex coronagraph imparts an azimuthal phase ramp to the wavefront.
\begin{equation}
    \widetilde{E}_{1, \text{post}}(\rho, \phi) = \text{exp}[im\phi]\cdot \widetilde{E}_{1, \text{pre}}(\rho, \phi),
    \label{eqn:E1_postVC}
\end{equation}
where $m$ is, as usual, the vortex topological charge. The Fourier transform of Eq. \ref{eqn:E1_postVC} yields the complex amplitude in the Lyot plane. 

\begin{equation}
    E_{2, \text{pre}}(r, \theta) = \iint \widetilde{E}_{1, \text{post}} \text{ exp}[i\rho r\text{cos}(\phi-\theta)]\rho d\rho d\phi
    \label{eqn:E2_preNRM}
\end{equation}
In this Lyot plane, we apply a non-redundant mask to create the interferometric aperture. 
We adopt the formalism of Tallon and Tallon-Bosc (1992)\cite{1992A&A...253..641T} to describe the NRM analytically. Let $\mathcal{E}_n(\vec r)$ be the transmission function of a single sub-aperture with number $n$, which is equal to zero outside of the sub-aperture and 1 within it. Each of the identical sub-apertures are centered upon their respective locations within the pupil, $\vec {r_n}$. The complex amplitude following the NRM is then:
\begin{equation}
    E_{2, \text{post}}(r, \theta) = E_{2, \text{pre}}(r, \theta)\cdot\sum_n  \mathcal{E}_n(\vec r - \vec {r_n}), 
    \label{eqn:E2_postVC}
\end{equation}
and the complex amplitude at the following focal plane becomes: 
\begin{equation}
    \widetilde{E}_3(\rho, \phi) = \text{FT}^{-1}\{E_{2, \text{post}}\}.
    \label{eqn:E3_detectorPlane}
\end{equation}
The intensity of the interferogram in this focal plane, $I_3 = |\widetilde{E}_3|^2$, is then measured by a detector, and the interferometry metrics are measured by taking another Fourier transform of the interferogram to the UV plane, which can be seen in the right column of Fig. \ref{fig:block_diagram}:
\begin{equation}
    \widetilde{U}_4(\vec k) = \text{FT}^{-1}\{I_{3}\},
    \label{eqn:general_UV_plane} 
\end{equation}
where $\widetilde{U}_4$ is the complex function in the UV plane resulting from the Fourier transform of the interferogram. $\vec k$ is the spatial coordinate of the UV plane, which is related to the spatial frequencies of the NRM baselines. 

The Fourier transform of an interferogram results in discrete "splodges" in the UV plane at spatial frequencies corresponding to the baselines of the NRM, as well as the zero-frequency coordinate. We describe the extraction of the visibility information from the UV plane in more depth in Sec. \ref{sec:observables}.

\subsubsection{Complex Amplitude at the NRM Input} \label{sec:ugly_math}
An interesting aspect of the vNRM arises 
when we derive analytic expressions for the Lyot plane for a telescope with a secondary mirror obscuration. From Fig. \ref{fig:block_diagram}, we use a pupil function of a circular, monolithic telescope aperture with radius $R_\text{P}$ and a circular secondary obscuration in the center with radius $R_\text{s}$. At this stage, 
we assume the static and dynamic, phase and amplitude aberrations are negligible for the following analytical development. Equation \ref{eqn:E0_entrance} then becomes: 
\begin{equation}
    \begin{split}
    \mathcal{P}(r, \theta) &= 
    \left\{
        \begin{array}{lr}
            0, & r > R_\text{P}\\
            1, &  r \leq R_\text{P}
        \end{array}
    \right\} - 
    \left\{
        \begin{array}{lr}
            0, & r > R_\text{s}\\
            1, & r \leq R_\text{s}
        \end{array}
    \right\} \\ 
    E_0(r, \theta) &= A_0 \cdot \mathcal{P},
    \end{split}
    \label{eqn:Circular_pupil}
\end{equation}
where $A_0$ is some arbitrary constant to denote the power in the wavefront.

The complex amplitude in the first focal plane, Eq. \ref{eqn:E1_preVC}, is the familiar expression:
\begin{equation}
    \begin{split}
    \widetilde{E}_{1, \text{pre}}(\rho, \phi) &= A_0*\left(\frac{R_\text{P}J_1(R_\text{P}\rho)}{\rho} - \frac{R_\text{s}J_1(R_\text{s}\rho)}{\rho}\right) \\
    \widetilde{E}_{1, \text{post}}(\rho, \phi) &= A_0*\left(\frac{R_\text{P}J_1(R_\text{P}\rho)}{\rho} - \frac{R_\text{s}J_1(R_\text{s}\rho)}{\rho}\right)\text{ exp}[im\phi],
    \end{split}
    \label{eqn:VC_plane}
\end{equation}
where $J_1(x)$ is the first-order Bessel function of the first kind.

We show the analytical propagation of $\widetilde{E}_{1,\text{ post}}$ to the Lyot plane in Appendix \ref{sec:math_appendix}. 
The classical vortex coronagraph result for an unobstructed circular pupil at the telescope entrance aperture yields zero complex amplitude inside the corresponding, re-imaged entrance aperture 
in the Lyot plane, having been optically filtered by the vortex\cite{2005ApJ...633.1191M, 2005OptL...30.3308F}. We refer to the re-imaged entrance aperture as the "Lyot region". 
However, as was shown in Ref. \citenum{2008MNRAS.384..515J} and App. \ref{sec:math_appendix}, the secondary obstruction injects light into the Lyot region. For a general even-charged vortex coronagraph, the complex amplitude in the Lyot region becomes: 
\begin{equation}
    \begin{split}
    E_{2, \text{pre}}(r, \theta) = A_0*2\pi\text{ exp}[im\theta]*
    \Bigg[&R_\text{P} \left\{
    \begin{array}{lr}
            0, & r < R_\text{P}\\
            \frac{R_\text{P}(m/2)!}{r^2((m/2) -1)!}F(\frac{m}{2}+1, 1-\frac{m}{2}; 2; \left(\frac{R_\text{P}}{r}\right)^2), &  r \geq R_\text{P}
        \end{array}
    \right\} \\
   &- R_\text{s} \left\{ 
    \begin{array}{lr}
            0, & r < R_\text{s}\\
            \frac{R_\text{s}(m/2)!}{r^2((m/2) -1)!}F(\frac{m}{2}+1, 1-\frac{m}{2}; 2; \left(\frac{R_\text{s}}{r}\right)^2), &  r \geq R_\text{s}
        \end{array}
    \right\}\Bigg] ,
    \end{split}
    \label{eqn:lyot_full}
\end{equation}
where $F$ is the Gaussian hypergeometric function, expanded for charges $m=2,4,6,\text{and } 8$ in Eq. \ref{eqn:expand_Ffunc} in Appendix \ref{sec:math_appendix}. 

By explicitly deriving the complex field in the Lyot plane, we have shown the complex amplitude to be spatially variable in the plane following the vortex coronagraph, which generates differences between the inputs of the vNRM and the standard NRM. 
When we then apply a non-redundant mask to the Lyot plane, the interferometric observables no longer directly measure the visibilities of the astrophysical source function, but the visibilities of the light that is filtered through the coronagraph. An important consequence of this for the vNRM methodology is that the observable products will now depend on the stability of the electric field in the Lyot region following the vortex coronagraph, which may be used to inform to estimate requirements on wavefront stability for the technique. 
Furthermore, the deviations from the standard NRM input may necessitate changes in the typical data reduction techniques for the interferometric observables. 
This will be discussed in more detail in section \ref{sec:Future_work_discuss}. 


\subsection{Details of the Numerical Investigation} \label{sec:sim_setup}
The propagation of off-axis light sources through a vortex coronagraph becomes complicated analytically\cite{2008MNRAS.384..515J}, as is the analytical description of an interferogram following an NRM. 
Instead, we now create a numerical simulation of the vNRM system to investigate its off-axis, high-contrast capabilities in HCIPy\cite{por2018hcipy}. In this section, we present the relevant parameters defining our simulation. 
As a motivating case, we reference an elementary rendition 
of the Habitable Worlds Observatory (HWO), as the goal of this observatory will ultimately be to directly detect and characterize exoplanets in the HZ of nearby Sun-like stars. 
HWO can provide context for making various reasonable design assumptions without limiting the scope of our investigation to any specific system architecture or instrumental performance. 

For the full telescope aperture, we simulate a 5\,m monolithic circular primary mirror with a 1\,m diameter secondary mirror obstruction --- ignoring any secondary support structures. 
We use an operating wavelength of 900\,nm, investigating a monochromatic portion of the \textit{z}-band. With this aperture size and wavelength, the angular resolution of the telescope would be $1.22\lambda/D \approx 45$\,mas. The target angular separation of 50\,mas would correspond to a separation of $1.35\lambda/D$. With a monochromatic assumption, we avoid simulating any chromatic effects between the depth of the vortex null and the breadth of the NRM fringes. While only simulating a single wavelength, for the purposes of defining a non-zero photon flux, we assume an effective spectral bandwidth of 0.2\%, which generates 
a photon flux of $4.8\times 10^9 \gamma/s$ for a magnitude $m_z=0$ star in the telescope entrance aperture. We conservatively assume a 5\% throughput through the instrument to account for additional losses due to beam splitters, reflections, adaptive optics, etc. This throughput does not account for losses intrinsic to the NRM or to the coronagraph. 
We use a super-Nyquist sampled imaging system, using 4 detector pixels per $\lambda/D$ resolution element in each dimension. With nearly 200 pixels to a side, the FOV of the detector spans approximately 3\,square arcsec, which focuses on the region near the IWA of the vortex. For the scope of this project, we only simulate the behavior of the system for a narrow FOV, to off-axis separations within a few $\lambda/D$. 


Vortex coronagraphs impart a helical phase shift to the wavefront. This is done using either a vector vortex, which utilizes a spatially varying polarization axis 
to impart the desired phase shift into the wavefront\cite{2005ApJ...633.1191M}, or a scalar vortex, which uses physical material to create an optical path difference that imparts the desired phase shift\cite{2019SPIE11117E..1FR}. To avoid making assumptions relating to polarization control and leakage, which can limit the performance of a vector vortex\cite{2010SPIE.7739E..14M}, we opt to simulate scalar vortices in this work.


A significant drawback of the scalar vortex is a strong chromatic leakage\cite{2019SPIE11117E..1FR}. 
While our monochromatic simulation will not be affected by this, in the context of HWO, the magnitude of on-axis starlight rejection will be required to be achromatic across astronomical passbands. There have been recent developments into improving the achromaticity of scalar vortices with a high degree of fidelity between simulations and testbed results\cite{2024JATIS..10a5001D, 2026arXiv260322537D, 2026arXiv260322551D}.
With this future goal in mind, we model the vortex coronagraph using the same wavefront propagation architecture as is used in the FALCO code to model the HCIT testbed\cite{2018SPIE10698E..2VR}. 
For now, we use a simple scalar sawtooth vortex for our coronagraph, equivalent to the one analytically described in Eq. \ref{eqn:E1_postVC}. We will primarily present the results from a charge 2 vortex coronagraph in the main body of this paper, but we repeat all simulations for charges 4 and 6 
and comment on differences in the results that arise from differing charges. 

As was derived in Sec. \ref{sec:ugly_math}, the presence of the secondary obstruction will limit the performance of the vortex coronagraph. While there are methods to improve the contrast achieved by a coronagraph in the presence of a secondary, we opt to only study the vNRM system with a standard vortex coronagraph with a central obstruction to simplify our simulation. 
In Sec. \ref{sec:noSec_discussion}, we repeat all tests of the vNRM system using an unobstructed entrance aperture, as 
removing the secondary mirror may provide some initial intuition towards the performance of the vNRM system when using a high-performance coronagraph.  

When combining the vortex with a NRM, there is a choice of which pupil plane to insert the NRM into. 
In the "vortex-first" (vNRM) configuration, the full aperture is brought to a focal plane with the coronagraph, and the NRM is inserted into the Lyot plane (as is depicted in Fig. \ref{fig:block_diagram}). In the "NRM-first" configuration, the NRM is inserted into the pupil prior to being brought to a focus. The interferogram in the following focal plane is centered on the vortex coronagraph, and then a second NRM with smaller subapertures is placed into the Lyot plane. 
The visibility metrics of the NRM can be measured for both of these configurations, but the throughput of the NRM-first configuration is quite low, leading to limited scientific use cases. 
As a result, we focus on the vNRM configuration in this work. We briefly address the NRM-first configuration throughput in Sec. \ref{sec:discussion}.

In this work, we use a simple 4-hole mask, shown in the aperture mask plane of Fig. \ref{fig:block_diagram}, for simplicity when interpreting and visualizing the simulated data. 
The locations of the subapertures in the NRM are listed in Table \ref{tbl:subap_loc}, and the subapertures have a diameter of $\sim0.6$\,m. 
With 4 sub-apertures, this NRM features 
6 unique baselines and 3 unique closure triangles. 
Many designs exist to utilize more sub-apertures for more thorough coverage of the UV plane or higher throughput\cite{2000PASP..112..555T, 2019AJ....157..249G, 2023PASP..135a5003S, 2023SPIE12680E..24L}, 
but limiting the number of probed baselines helps to visualize the signal present in each individual baseline and closure phase triangle and the impact of the vortex on each of those signals. In the next section, we elaborate on the extraction of the visibility metrics and compare the signal from a high-contrast companion between the standard NRM and vNRM configurations. 
\begin{table}[h]
    \centering
    \begin{tabular}{|c|c|}
        \hline
        Subaperture & Location  \\
        Index & (\textit{X}, \textit{Y}) (m)   \\
        \hline 
        \rule[-1ex]{0pt}{3.5ex}  0 & (-1.1, 1.375)     \\
        \hline
        \rule[-1ex]{0pt}{0ex}  1 & (-1.1, 0)   \\
        \hline
        \rule[-1ex]{0pt}{0ex}  2 & (1.1, 0)    \\
        \hline
        \rule[-1ex]{0pt}{0ex}  3 & (0, -1.65)    \\
        \hline
    \end{tabular}
    \vspace{3pt}
    \caption{NRM subaperture locations.}
    \label{tbl:subap_loc}
\end{table}


\subsection{Interferometric Observables with a NRM and Vortex} \label{sec:observables}

In this work, we use the squared visibilities and closure phase angles to probe the visibility information corresponding to the astrophysical source. 
Classically, the visibility information from an NRM interferometer is present in the UV plane as a convolution between the complex visibilities, 
located at their respective baselines, and the optical transfer function of the sub-aperture. We denote the complex visibility of the fringe between subaperture $i$ and subaperture $j$ as $\widetilde{V}_{ij}$. 
The complex visibilities are typically recovered from the UV plane splodges by using a Hanning window function to take weighted samples of the centers of each splodge at the expected location of the spatial frequency of the baseline\cite{1999PhDT........19M, 2000PASP..112..555T, 2017ApJS..233....9S, 2019AJ....157..249G}. 
We follow this convention by convolving the window function with $\widetilde{U}_4(\vec k)$, normalizing by the complex value at the origin, and extracting the complex visibilities from their baseline locations. 
Then, the squared visibilities and closure phases are defined as, 
\begin{equation}
    \begin{split}
    |V_{ij}|^2 &= \left|\widetilde{V}_{ij}\right|^2  , \text{ and} \\
    \text{CP}_{ijk} &= \text{arg}\left(\widetilde{V}_{ij} \right) + \text{arg}\left(\widetilde{V}_{ki} \right) - \text{arg}\left(\widetilde{V}_{kj} \right), 
    \end{split}
    \label{eqn:Signal_definition}
\end{equation}
where $\text{CP}_{ijk}$ denotes the closure phase between apertures $i$, $j$, and $k$, and $\text{arg}(\text{ })$ returns the argument of a complex number.

For the purposes of this study, the only astrophysical source we consider is an on-axis point source target with a fainter, off-axis companion point source. 
While disks and other extended sources are of astrophysical interest at small angular scales, we do not yet consider how to simulate, reconstruct, or evaluate observations of these sources. 
Instead, we characterize the behavior of the vNRM system by investigating the response of the interferometric observables, defined by Eq. \ref{eqn:Signal_definition}), 
to companion point sources at different radial separations, azimuths, and contrasts to the on-axis source. We define the companion signal in the observables as those measured for an on-axis target with a companion subtracted by the observables measured on a reference on-axis target, which mirrors the strategy that would be used in an on-sky observation. 

\begin{figure}[t] 
\centering
\includegraphics[width = 0.98\textwidth]{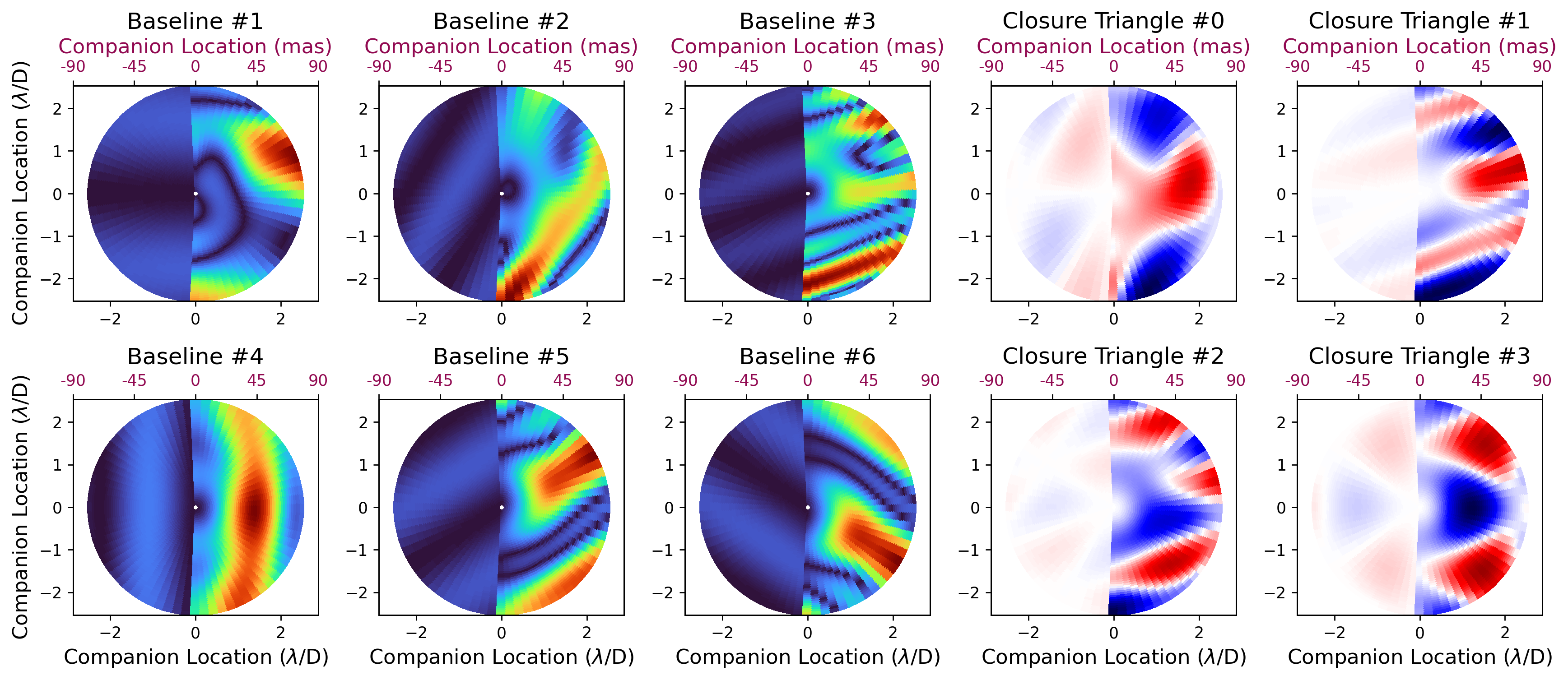}
\caption{Comparison of the companion signal in interferometric observables between the standard NRM (left halves) and charge-2 vNRM configurations (right halves).} \label{fig:NRMANDvNRM2_pureSignal} 
\end{figure}

In Fig. \ref{fig:NRMANDvNRM2_pureSignal}, we show an example of the variations in the companion signal 
for both the standard NRM (left halves) and the vNRM (right halves) in each of the baselines and closure phase triangles in response to a high-contrast companion at varying field locations. 
This is created in our simulation by incoherently summing the interferogram of an on-axis wavefront and the interferogram of a series of wavefronts at pre-determined off-axis positions and contrast ratios. We then extract the companion signal for each location. 
The differences in closure phases are simply reported in degrees while the differences in squared visibilities are normalized by their references to show the fractional signal change.

The responses of the standard NRM squared visibilities to a companion have a sinusoidal spatial dependence, which is expected as this observable probes the sinusoidal, mutual-intensity fringe response that each two-aperture baseline projects onto the sky. 
Comparing the signals of the NRM and vNRM systems in response to a $10^{-3}$ companion, as is depicted in Fig. \ref{fig:NRMANDvNRM2_pureSignal}, it is clear that adding a vortex coronagraph increases the magnitude of the interferometric signal that can be detected, improving the sensitivity by approximately an order of magnitude. 
It can also be visually confirmed that the spatial response of the vNRM is similar to the NRM, building confidence that the vNRM observables retain information about the astrophysical scene following the vortex coronagraph. 
There still remains a non-linear transformation between the signal of the NRM and the signal of the vNRM, so while data reduction techniques for NRM data exist\cite{1999PhDT........19M, 2017ApJS..233....9S, 2019AJ....157..249G}, the adaptation of these pipelines for the vNRM beyond the scope of this paper. We elaborate more on this topic in Sec. 
\ref{sec:Future_work_discuss}. 


We refer to the data represented by Fig. \ref{fig:NRMANDvNRM2_pureSignal} as "pure signal" because its creation did not include any potential sources of random or systematic error beyond machine error. 
In the following sections, we detail and conduct numerical experiments to investigate the properties of the measurements from the vNRM and NRM in applicable astronomical observing contexts. 
In Sec. \ref{sec:Photon_noise}, we investigate the sensitivity of the visibility metrics to photon noise, and in Sec. \ref{sec:WFE_general}, we investigate the robustness of the NRM and vNRM signals in the presence of static and dynamic wavefront aberrations. 

\section{Numerical characterization of photon noise sensitivity} \label{sec:Photon_noise}
To build estimations of potential instrument performance, we must consider sources of noise that could be mistaken for signal. 
As opposed to traditional imaging, where the observable signal increases as more photons are collected, the signal in the squared visibilities and closure phases depends purely on the parameters of the astrophysical source that is being observed. In this case, changing 
the number of collected photons only affects the photon noise uncertainty of the measurement. 


As an implicit drawback to the use of an aperture mask, the throughput of a NRM system limits the instrument using the system to brighter targets. 
In this section, we quantify the photon limit of the vNRM system and compare to an equivalent standard NRM setup. We assume that these simulated observations are photon-noise limited (i.e. negligible contributions from detector noise or wavefront error) to illustrate the "best-case" scenario for companion detection with the vNRM. 

\subsection{Methods for Estimating the Photon Noise Limit} \label{sec:photon_noise_snr_methods}
To calculate the photon noise limit of the vNRM system, we quantify the influence that photon noise at the detector has on 
the measurement of the visibility metrics in the UV plane. 
Photon noise depends on the number of photons collected by the detector in an exposure time. As a result, in quantifying photon noise, there is a degeneracy between the instrument throughput, exposure time, target brightness, etc. 
We break this degeneracy by using the photon arrival rate, defined by the simulation assumptions in Sec. \ref{sec:sim_setup}, for a $m_z = 5$ star, such that the number of photons expected to arrive on the detector 
is determined solely by the integration time. We investigate the photon limit of a $m_z = 5$ star because this is the faint end of the proposed target list for HWO as of the time of writing\cite{2024AJ....167..139T}.

We investigate the impact of photon noise on the visibility measurements for a series of integration times between 0.01 and 500 seconds. 
When noise is present in the interferogram, 
its projection to the UV plane will include deviations in the amplitude and phase of the complex visibility splodges away from their reference values. 
We generate a series of simulated photon distributions from the noiseless reference interferogram. After projecting each noisy interferogram to the UV plane, the photon noise can be found as the standard deviation in the resulting extracted visibility metrics. The amount of photon noise for the vNRM and NRM systems are then recorded as functions of the simulated integration time for the individual baselines and closure phase triangles. 

With the estimates of the photon noise in each visibility metric, we can estimate the signal to noise ratio (SNR) of observations of an off-axis companion. 
By comparing the magnitude of the signal in the squared visibilities and closure phases generated by a target with a high-contrast companion to the magnitude of the photon noise from the observation of a single on-axis source, we arrive at the SNR of measuring a given companion in one of the baselines or closure phase triangles at the photon noise limit: 
\begin{equation}
    SNR_\gamma = \frac{S}{\sigma_\gamma},
    \label{eqn:SNR_photonNoise}
\end{equation}
where $S$ is the signal, calculated from Eq. \ref{eqn:Signal_definition}, and $\sigma_\gamma$ is the calculated photon noise. 
This is related to the likelihood of making a false-positive detection caused by photon noise. 
In Sec. \ref{sec:photon_snr_results}, we compare results from the NRM and vNRM systems for the photon SNR limit in each baseline and closure phase triangle detecting an off-axis companion.

Another approach to represent the photon limit of the NRM and vNRM systems is to present this data as a contrast curve
-- the highest contrast object at a given off-axis separation that could be detected to some target level of confidence. We do this by utilizing the fact that the visibility signals scale with the flux of the off-axis high-contrast companion, $S\approx s'\cdot 10^{\text{contrast}}$, where $s'$ is some baseline companion signal value and "$\text{contrast}$" is the companion contrast. The SNR equation, thus, can be converted to give the contrast as a function of the other parameters. We choose to implement a target SNR of $SNR_\gamma =5$. While it will be necessary to fully consider the noise properties and sampling statistics of the 
detection method to verify what SNR is necessary to make a $5\sigma$ detection of a physical companion\cite{2014ApJ...792...97M}, $SNR_\gamma =5$ is a straightforward initial estimation. The equation to estimate the contrast curves for the NRM and vNRM systems for different magnitudes of photon noise is then 
\begin{equation}
    C \approx \text{log}_{10}\left( \frac{\sigma_\gamma \cdot SNR_\gamma}{s'} \right).
    \label{eqn:PhotonNoise_CC}
\end{equation}
The expected contrast curves are also plotted in Sec. \ref{sec:photon_snr_results}.

\begin{figure}[t] 
\centering
\includegraphics[width = 0.95\textwidth]{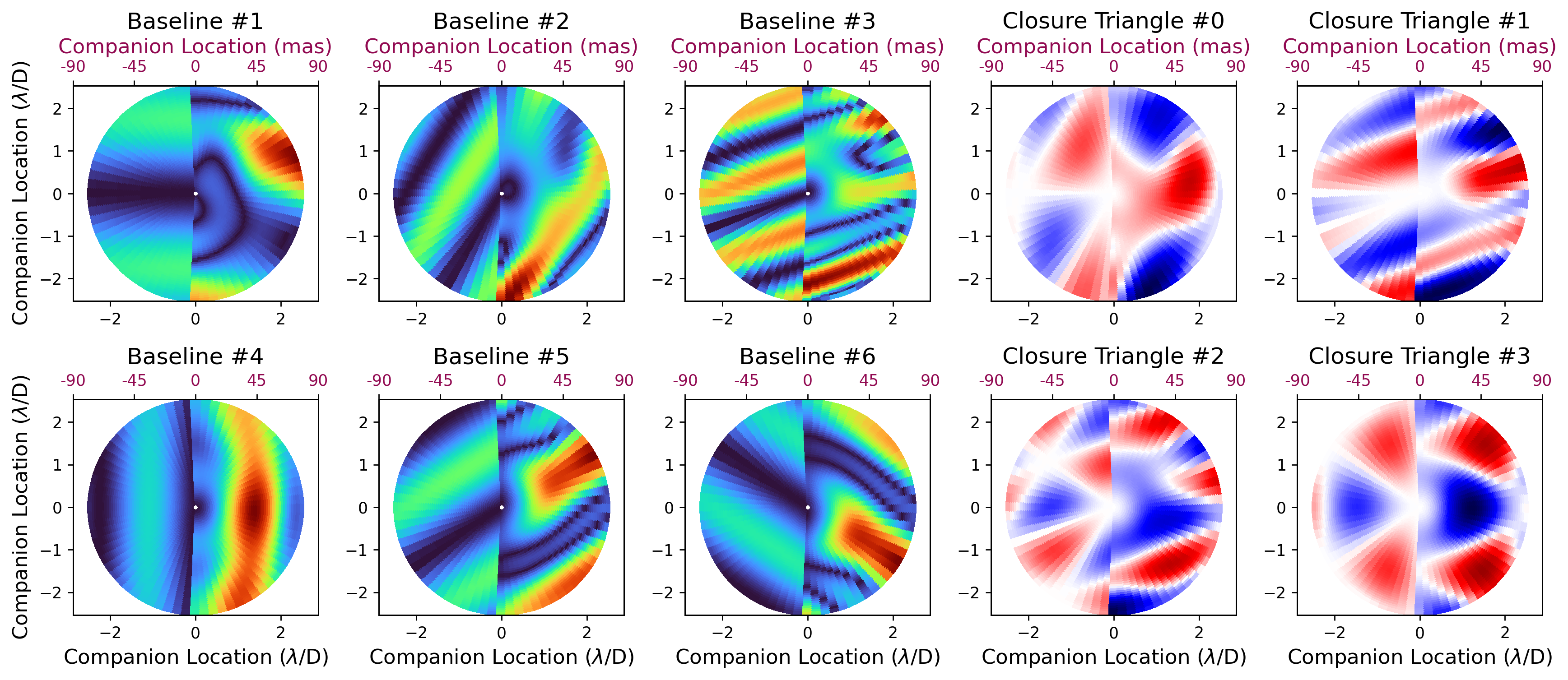}
\caption{Comparison of the photon-limited SNR maps of measurements of the interferometric observables to an off-axis companion. As was done in Fig. \ref{fig:NRMANDvNRM2_pureSignal}, we compare the standard NRM arrangement (left) and a charge-2 vNRM arrangement (right).} \label{fig:SNRmap_NRMANDvNRM2} 
\end{figure}

\subsection{The Photon Noise Limit} \label{sec:photon_snr_results}

By comparing the interferometric observable signals to the photon noise in the manner described in the previous section, we calculate estimations of the photon limit of the NRM and vNRM systems. In Fig. \ref{fig:SNRmap_NRMANDvNRM2}, we compare the expected SNR of the NRM and vNRM visibility measurements in response to an off-axis companion as a function of the companion's simulated position. In this figure, we simulated a $10^{-3}$ companion to the on-axis $m_z=5$ target with 10\,seconds of integration time. The main takeaway from this figure is that the estimated SNR of an observation with the vNRM is higher than the SNR of an equivalent observation with a standard NRM. In every baseline and closure phase triangle, the peak estimated SNR value in the vNRM maps are higher than the corresponding peaks in the NRM maps. In the context of this specific experiment, this would mean that it would take less integration time to reach a target SNR in a measurement by using a vNRM compared to a standard NRM in the photon limited regime.

\begin{figure}[t] 
\centering
\includegraphics[width = 0.93\textwidth]{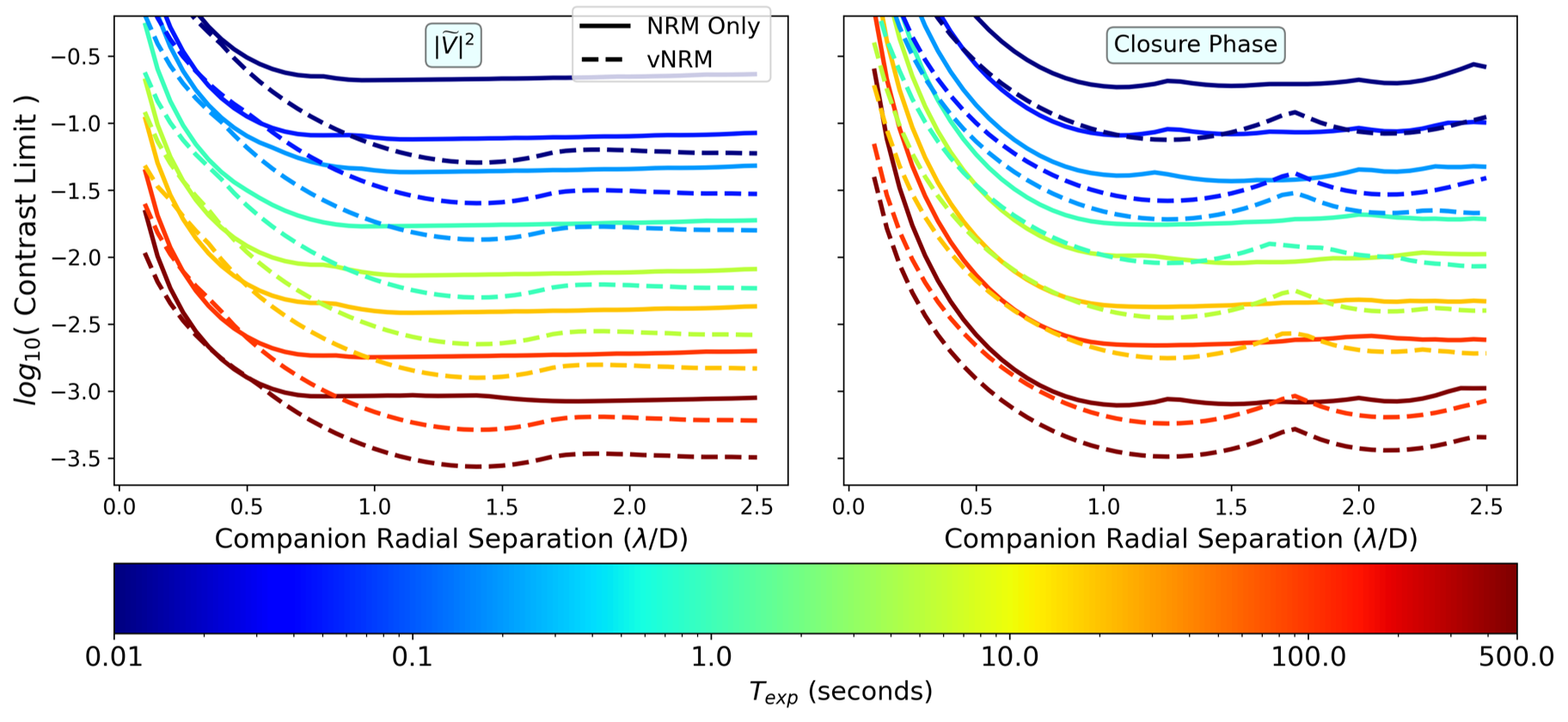}
\caption{Comparison of the contrast curves of the interferometric observables for different exposure times of photon noise-limited integrations. Contrast curves show the faintest companions that could be detected with a $SNR=5$ measurement in the exposure time. This figure separates the contrast curves for the squared visibilities (left) and the contrast curves for the closure phases (right).} \label{fig:PhotonNoise_CC_c2_w2nd} 
\end{figure}

From Eq. \ref{eqn:PhotonNoise_CC}, we can turn these SNR maps into the contrast limit for an $SNR=5$ measurement as a function of the companion location and exposure time for each baseline and closure phase triangle. To present this information as a typical contrast curve, we will want to condense the azimuthal dimension so that the contrast depends purely on the radial separation. The azimuthal response of the interferometric signals is dependent on the baseline and closure triangle choice. As we selected only a simple, illustrative NRM design, we assume that, by optimizing for UV coverage and throughput, 
we could develop a mask with a typical discovery space that is consistently similar to the best performing azimuthal location at each radial separation delivered by this simple mask. 
Thus, to show the faintest off-axis companions that can be detected, we take the minimum value of the contrast limit in the azimuthal dimension as well as across the baselines and closure triangles. In Fig \ref{fig:PhotonNoise_CC_c2_w2nd}, we show the contrast curves for the NRM (solid lines) and vNRM (dashed lines) for given exposure times. 
This figure shows the contrast curves with the charge-2 vortex. 

We can quantify the improvement in performance between the NRM and vNRM systems by taking the difference between the contrast curves for both systems at each integration time. 
For the charge-2 vortex, the squared visibility contrast curves 
from the vNRM reach about 0.5 orders of magnitude deeper than the corresponding standard NRM curves for companions at separations of $\geq 1\lambda/D$ and reach comparable contrasts within $0.5\lambda/D$ separations. The closure phase contrast curves are a similar 0.5 orders of magnitude deeper for the vNRM compared to the standard NRM but maintain this improvement for all companion separations. For a charge-4 vortex, the contrast curve improvement from vNRM to NRM is 0.4 orders of magnitude for both squared visibilities and closure phases with the improvement manifesting at companion separations of $\gtrsim 1.3\lambda/D$. For a charge-6 vortex, the improvement is slightly less than charge-4 with the improvement manifesting at separations of $\gtrsim 1.6\lambda/D$. This aligns with expectations that higher-charge vortices have larger inner-working angles and suppress more off-axis light. The magnitude of contrast gain is largely determined by the raw contrast achieved by the coronagraph. In Sec. \ref{sec:noSec_discussion}, we repeat this experiment with a "high-performance" coronagraph setup that achieves a higher initial contrast gain, 
and we show how this different architecture reflects in deeper contrast curves for the vNRM.

\section{Numerical Characterization of Wavefront Error Sensitivity} \label{sec:WFE_general}
In addition to photon noise, a major limitation to high-contrast observations is the error and uncertainty induced by wavefront error (WFE). While some visibility metrics of NRM observations can be self-calibrated against WFE\cite{1958MNRAS.118..276J}, 
the addition of a vortex coronagraph creates a non-uniform electric field in its subsequent Lyot plane, which is dependent on the WFE at the entrance aperture\cite{2008MNRAS.384..515J}. 
These non-uniformities result in changes to the interferogram and visibility metrics, which can be mistaken for signal from an astrophysical source and limit our confidence in the accuracy of the recovered visibilities. 

In this paper, we separate WFE into two types: static WFE and dynamic WFE. Static WFE refers to WFE that accrues in the instrument between observation of a reference source and the scientific observation, but remains static over the course of a single exposure. In the high-Strehl regime, where a wavefront is being actively corrected to minimize phase aberrations, 
static WFE will cause modulations in the fringe position and shape inside the interferogram. The static WFE thus causes changes in the visibility metrics, but this WFE can be mitigated through higher calibration cadences or potentially through post-processing. 
Dyanmic WFE refers to the evolution in wavefront phase that occurs during a detector exposure, which will cause the fringes in the interferogram to broaden and smear together. Dynamic WFE will cause irrecoverable loss in the visibility measurement. 
In this section, we quantify the contrast limit of the vNRM system in response to wavefront error. We assume that this contrast limit will reflect observations with the system when it is WFE limited (i.e. sufficient exposure time to achieve comparably negligible photon noise), thus, revealing further insight into the expected performance from the vNRM. 

%

\subsection{Methods for Estimating the WFE Limit} \label{sec:methods_wfe}
We take a similar approach for quantifying the effect of WFE on the visibility measurements as we did with photon noise. Random WFE will create random variations in the visibilities, so we quantify the uncertainty in the visibility metrics as a function of the amplitude and mode of the WFE. We define phase aberrations in the Fourier mode basis for the purpose of distinguishing modes by spatial frequency. 
The Fourier modes are created 
as 2D sinusoids with spatial frequencies of an integer number of cycles across the telescope aperture. 
Aside from the frequency, the arguments of a sinusoid function are its amplitude and phase. 


Our procedure is to induce a WFE of a specific spatial frequency and amplitude in the entrance aperture and then measure the resulting deviation of the interferometric observables. 
The resulting error in the visibility metrics is averaged across the phase of the Fourier modes, but otherwise quantified in the same manner as the photon noise in Sec. \ref{sec:Photon_noise}. It will become clear that the sensitivity of the NRM and vNRM is generally consistent between WFE of low-order spatial frequencies and of high-order spatial frequencies. Thus, we simplify the investigation by exploring a set of random combinations of low-order WFE and of high-order WFE at a series of amplitudes and reporting the SNR of companion detections as a function of these WFE amplitudes. As with the photon noise limit, we will be able to invert the SNR estimates of each visibility metric in response to companions of variable contrast to also estimate the contrast curve of the WFE-limited observations. The results of this investigation are reported in Sec. \ref{sec:results_wfe}.

The sensitivity of the vNRM to static WFE is found by taking a series of random linear combinations of all of the low-order or high-order Fourier modes at random amplitudes and phases, normalizing the resulting static WFE to some target rms amplitude across the entrance aperture, and then measuring the standard deviation of the visibility metrics across these different realizations of the random WFEs. This process is repeated several dozen times for each amplitude to understand the repeatable impact of static uncontrolled WFE at low and high-order spatial frequencies. 

The characterization of the vNRM sensitivity to dynamic WFE is more complicated than static WFE due to the larger domain of degrees of freedom. While the most applicable result would show the contrast limit of the vNRM as a function of the residual WFE following an adaptive optics (AO) system, challenges in the simulation of realistic atmospheric wavefront evolution\cite{2000JOSAA..17.1650S, 2007MNRAS.378.1177B, 2014MNRAS.440.1925G, 2022SPIE12185E..83C} would limit the validity of those results. 
Instead, we assume some "steady-state" dynamic WFE behavior where the amplitude of each Fourier mode is equally, randomly unconstrained, and the total phase aberration is normalized to the target WFE, and we sum the resultant interferogram for several concurrent "time steps". Between each time step, we linearly progress the phase of each Fourier mode while applying a new random amplitude. The motivation for this is to approximate AO residuals by incoherently summing correlated dynamic WFE with temporally-varying amplitude over an exposure. Ultimately, the difficulties in simulating residual dynamic WFE may limit the applicability of our findings to physical systems. %

\begin{figure}[t] 
\centering
\includegraphics[width = 0.87\textwidth]{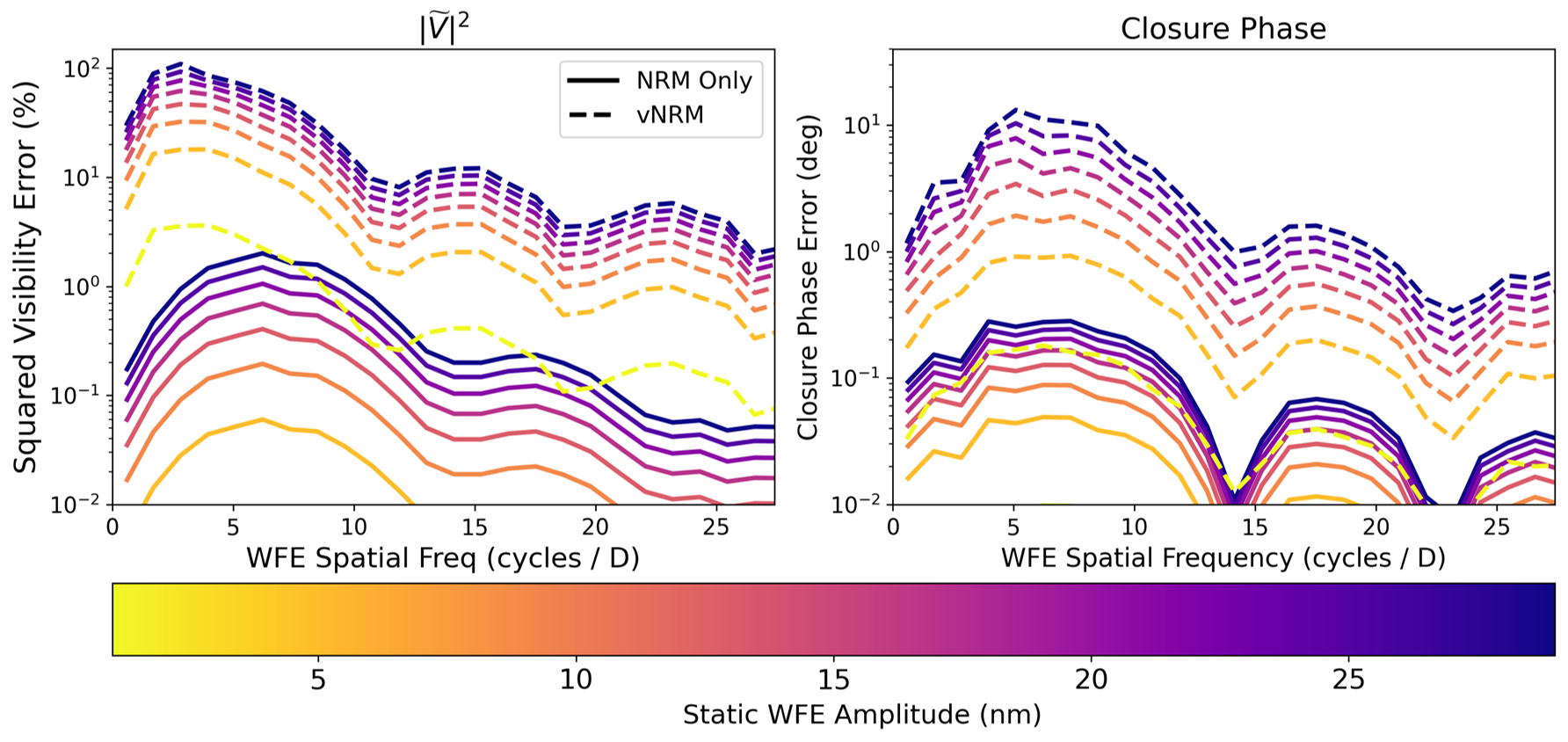}
\caption{Uncertainty induced into the visibility metrics by static WFE as a function of the spatial frequency of the WFE. This figure separates the WFE sensitivity for the squared visibilities (left) and the WFE sensitivity for the closure phases (right) for a standard NRM and charge-2 vNRM.} \label{fig:WFE_sensitivity} 
\end{figure}

\subsection{The Wavefront Error Limit} \label{sec:results_wfe}
The addition of a vortex coronagraph to an NRM system means that the interferogram will no longer be translationally invariant as the post-coronagraph PSF will depend on the position of the on-axis source with respect to the coronagraph. It, then, comes as no surprise that the observables of the vNRM are more sensitive to WFE than the NRM. Here, we quantify this sensitivity for both static and dynamic WFE. 

In Fig. \ref{fig:WFE_sensitivity}, we show the error 
in the visibility metrics in response to static WFE as a function of the WFE spatial frequency. From this, we can see that the vNRM has a higher WFE sensitivity than the NRM. We can also see that there is a clear difference between the response to low-order WFE and to high-order WFE. The cutoff between low-order and high-order corresponds to the spatial frequency of two cycles across the subaperture. At this spatial frequency, the artifacts generated in the focal plane are no longer co-located with the central interferogram and do not significantly influence the measurements extracted from the UV plane. 
The sensitivity curve for dynamic WFE has a similar spatial frequency dependence as Fig. \ref{fig:WFE_sensitivity}. 
The squared visibilities for the standard NRM are an order of magnitude more sensitive to dynamic WFE than static WFE, while the vNRM is approximately a factor of 2 \textit{less} sensitive to dynamic WFE in both the squared visibilities and closure phases. The charge-4 vNRM has similar sensitivity to WFE as the charge-2 vNRM, while the charge-6 vNRM experiences significantly more measurement error in response to WFE than the charge-2 or charge-4 vNRM. 

\begin{figure}[t] 
\centering
\includegraphics[width = 0.93\textwidth]{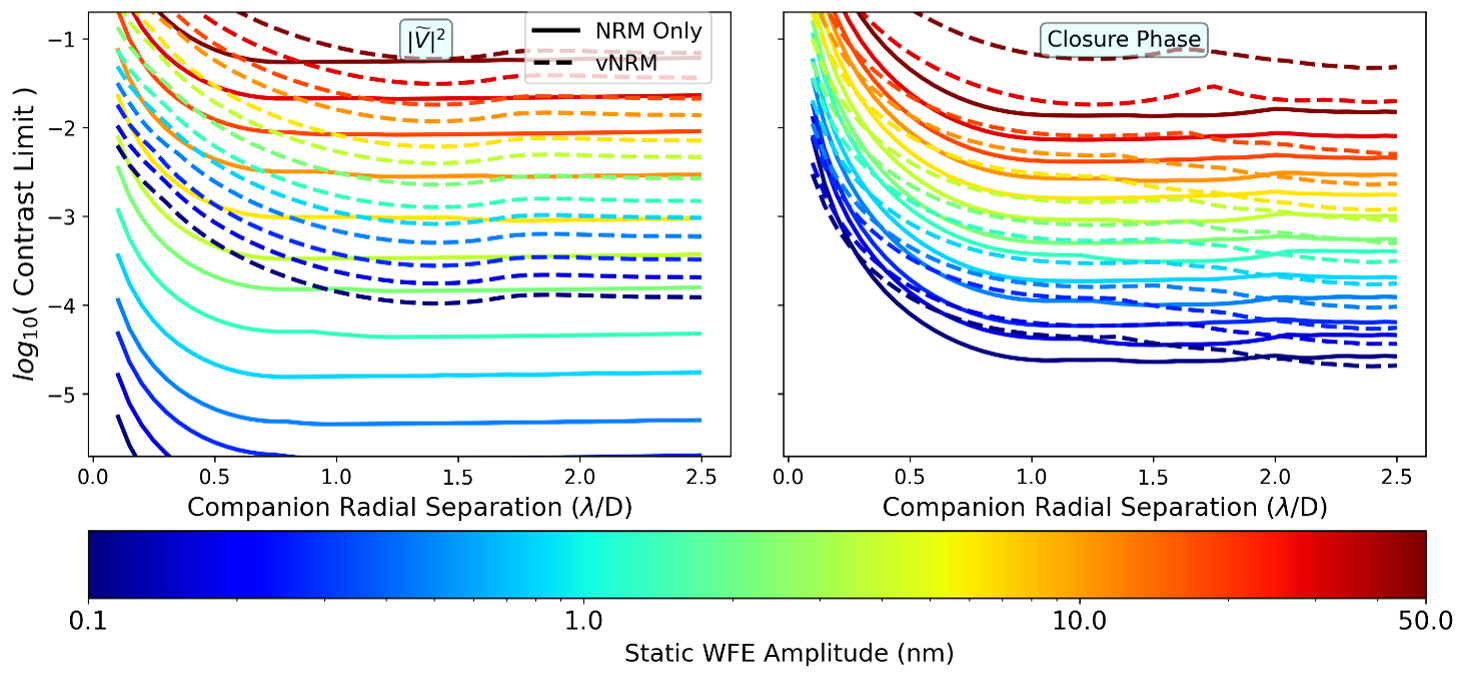}
\caption{Contrast curves for WFE-limited observations, presented in the same manner as Fig. \ref{fig:PhotonNoise_CC_c2_w2nd}. This figure separates the contrast curves for the squared visibilities (left) and for the closure phases (right) for a standard NRM and charge-2 vNRM.} \label{fig:WFE_CC_static} 
\end{figure}

In Fig. \ref{fig:WFE_CC_static}, we show the WFE-limited contrast curves for both the NRM and vNRM in response to static low-order WFE. In the closure phases, we see that, despite the increased sensitivity to WFE, the contrast limit of the vNRM is very similar to that of the standard NRM. At 50\,nm rms of low-order static WFE --- which corresponds to 50\,nm rms of change in the uncorrected WFE between a reference target and the science target --- high-confidence detections with both the squared visibilities and closure phases from the vNRM would be limited to contrasts of 0.1 or brighter. This corresponds to a similar limiting contrast of 0.1 for the NRM squared visibilities and $2\times10^{-2}$ for its closure phases. The contrast limit of the vNRM closure phase and the squared visibilities of both the vNRM and NRM improve by roughly an order of magnitude of contrast for each order of magnitude of improvement in WFE control. 
The contrast curve for the NRM squared visibilities improves at ten-fold this rate, achieving contrasts two orders of magnitude deeper for each order of magnitude of WFE amplitude reduction. 
The discrepancy between the contrast curves can be explained because the squared visibilities of the standard NRM are limited by speckle intensities in the final focal plane, which scale with the square of wavefront error, while the squared visibilities of the vNRM are constrained by the amplitude of the electric field leaking through the vortex. The closure phases are affected by interference between fringes, which scales with the amplitude of the focal plane electric field. These all scale linearly with the wavefront error in the entrance pupil plane, as opposed to quadratically. 
The contrast curves for the high-order WFE are approximately an order of magnitude deeper than the low-order WFE and follow the same scaling relationships.

The contrast curves for the dynamic WFE are approximately half of an order of magnitude deeper than their corresponding static WFE contrast curves. 
We provide context for these contrast curves and compare them with the photon noise contrast curves in Sec. \ref{sec:results_discussion}.

\section{Discussion} \label{sec:discussion}


In this discussion, we will apply our results to two realistic observing scenarios: space-based observations and an extreme adaptive optics (XAO)-fed instrument. We also briefly investigate the vNRM behavior when using a higher-performance coronagraph,  
and consider 
future directions for continuing development of the vNRM concept. 

Prior to these discussion topics, it is necessary to briefly justify the "vortex-first" architecture, rather than the "NRM-first" architecture for the vNRM system. 
The NRM separates the aperture into distinct locally-unobstructed pupils. The complex amplitudes from these pupils in the subsequent focal plane --- though interfering with each other --- are affected by a vortex mask in the same manner as was described in Sec. \ref{sec:math}, which leads to the on-axis light being entirely suppressed by a "Lyot-NRM" in the following pupil plane. While this would appear to be a desirable result, the drawbacks of this architecture become evident when considering off-axis performance. 
A companion located at a separation of $1\,\lambda/D$ from its on-axis host in the PSF of the full aperture would only be separated by $\frac{d}{D}\,\lambda/d$ in the PSF of a single subaperture. In our simulation, this ratio of the sub-aperture to full aperture diameter is $\frac{d}{D}=0.12$, which is a larger aperture ratio than is typical in existing NRM systems. 
At these separations, the coronagraph will significantly suppress the light from close-separation companions. The off-axis throughput and achievable contrast for this architecture improves for separations on the scale of $\frac{D}{d}\,\lambda/D$, but these spatial scales are already accessible to traditional high-contrast imaging techniques. 
Lacking a motivating science case, we did not investigate the sensitivity to photon noise or WFE for this architecture. 

\subsection{Providing Context for the Contrast Limits} \label{sec:results_discussion}
In Sec. \ref{sec:results_wfe}, we presented contrast curves that inform the potential performance limits of a vNRM limit in the presence of static and dynamic WFE. The static WFE contrast curves can be used to inform necessary stability requirements of phase aberrations and non-common path aberrations (NCPAs) over long timescales. The dynamic WFE contrast curves reflect requirements on AO performance and short-term stability. In Sec. \ref{sec:photon_snr_results}, we presented contrast curves for observations that were limited by photon noise, rather than WFE. There is an inherent tradeoff between the integration time and the amount of WFE that accrues during the observation. In this section, we explore this tradeoff, give context to these contrast levels, and allow that to inform the applicability of this technique into different observing conditions. 

At an operating wavelength of 900\,nm, our upper bound WFE of 50\,nm rms corresponds to a Strehl ratio of $88.5\%$, which is comparable to the achievable Strehl ratios of many of the best performing XAO systems in the field as of writing \cite{2013ApJ...776..130D, 2020SPIE11448E..0WM, 2024arXiv240713019K, 2024arXiv240719188L}. 
At these Strehl ratios, it can be asserted that the residual WFE is not dominated by low-order static WFE, but is instead distributed between dynamic WFE and high-order static WFE. 
However, it's difficult to make specific estimates without access to the telemetry data of an AO system. 
In the case that the 50\,nm of rms WFE of an XAO system is split evenly between dynamic and high-order static WFE (29\,nm rms each), the uncertainty from the corresponding error would limit $SNR=5$ detections to $6\times10^{-3}$ contrast companions in a single detection. This contrast limit corresponds to the photon limit at an exposure time of approximately 5\,sec. 
These coinciding contrast limits inform both limits on the decorrelation timescale for the residual wavefront phase, and also the maximum necessary exposure time, which allows for as many independent observations as possible. 
In the case that the vNRM system is limited by dynamic WFE, the contrast limit can be decreased with the square root of the number of repeated exposures, as the dynamic WFE in one exposure will be independent of the WFE in subsequent exposures. More work will need to be done to thoroughly understand the evolution of quasi-static WFE and their impact on the contrast limit in the case that the vNRM system is limited by static WFE. 

In the context of a space-based observation, we can assume zero induced WFE from atmospheric turbulence, so the contrast curve of an observation will be determined purely by the instrument stability and integration time. 
Analysis of the optical stability of the JWST estimated that the typical high-frequency, tip-tilt jitter to be at the level of approximately $\sim 13$\,nm and the worst-case drift in rms optical path due to thermal changes to be $\leq20$\,nm over hundreds of hours following a change in telescope pointing\cite{2022SPIE12180E..0VK}. Future space-based observatories (e.g. Roman, HWO, etc) intend to achieve further improved stability to the level of $<1$\,nm\cite{2025SPIE13623E..0CE}. JWST-levels of dynamic stability would correspond to a contrast limit of $6\times 10^{-3}$ in a single exposure, which can be reached in the photon limit in 10\,sec of integration. In the Roman-level stability case, the contrast limit is $\leq5 \times 10^{-4}$ in  a single exposure, corresponding to an exposure time of 10\,minutes to reach the photon limit. These exposure times are much less than the characteristic timescale for worst-case thermal drift, so uncertainties in measurements can be decreased further with repeated independent exposures. 

\subsection{Performance with an off-axis telescope} \label{sec:noSec_discussion}
The results from sections \ref{sec:photon_snr_results} and \ref{sec:results_wfe} showed how adding a vortex coronagraph to an NRM system can improve the photon-limited contrast curve of the instrument at the cost of increased sensitivity to wavefront error. 
We assert that the improvement in the vNRM contrast limit 
is constrained by the raw contrast achieved by the coronagraph. As is well understood, the characteristics of the secondary obstruction are important to the amount of starlight suppression a vortex coronagraph achieves\cite{2008MNRAS.384..515J, 2022A&A...665A.136P}, so the results found in sections \ref{sec:photon_snr_results} and \ref{sec:results_wfe} may be limited by an unideal choice in coronagraph. 
In this section, we briefly investigate how the characteristics of a vNRM would change when operating with a higher-performance coronagraph. 

Realistically, there are several options for improving coronagraph performance in the presence of a secondary mirror. Amplitude-apodized coronagraphs\cite{2017SPIE10400E..0OE}, phase-induced amplitude apodizing (PIAA) coronagraphs \cite{2018SPIE10703E..59L, 2026A&A...708A.144T}, and tandem vortex coronagraphs \cite{2011OptL...36.1506M, 2016OptCo.379...64S} have all shown improved performance of on-axis light suppression with a small inner-working angle (IWA). Each of these approaches could be used to achieve a high-performance coronagraph to be combined with a NRM system, which we speculate on more in Sec. \ref{sec:Future_work_discuss}. However, simulating these is beyond the scope of this project. 
Instead, to estimate the behavior of a high-performance coronagraph leading into an NRM, we repeat the experiments of sections \ref{sec:Photon_noise} and \ref{sec:WFE_general} for a charge-2 vortex coronagraph with the unobstructed entrance aperture, which could reflect the vNRM performance when used on an off-axis telescope. 

\begin{figure}[t] 
\centering
\includegraphics[width = 0.83\textwidth]{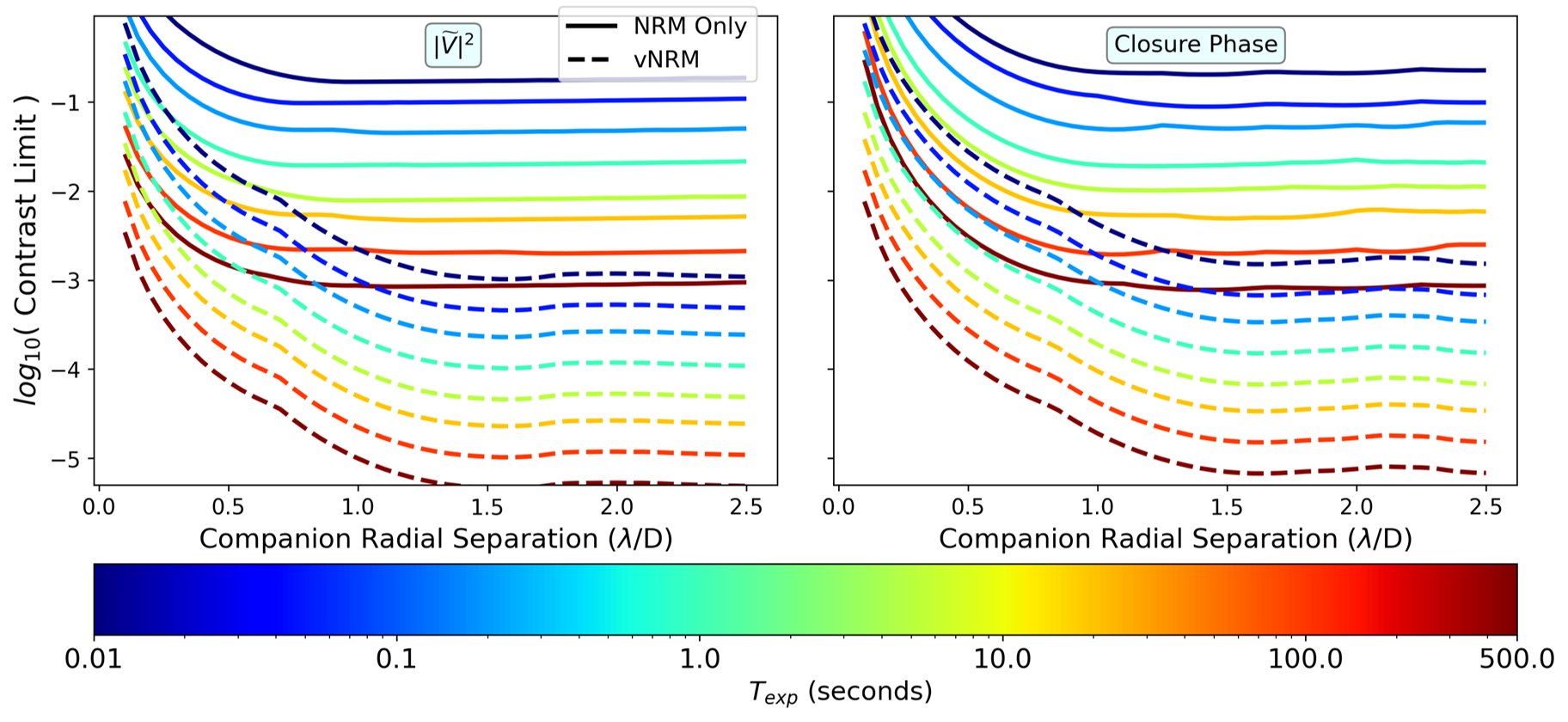}
\caption{Equivalent figure to Fig. \ref{fig:PhotonNoise_CC_c2_w2nd} for a vNRM on an off-axis telescope. Shows the photon-limited $SNR=5$ contrast curves for squared visibilities (left) and closure phases (right).} \label{fig:PhotonNoise_CC_c2_No2nd} 
\end{figure}

In Fig. \ref{fig:PhotonNoise_CC_c2_No2nd}, we show the photon-limited contrast curves for the high performance vNRM. In comparison to Fig. \ref{fig:PhotonNoise_CC_c2_w2nd}, the high performance vNRM detects significantly higher contrast companions to a confidence of $SNR=5$ for the same integration time. In this case, adding the coronagraph improves the photon-limited contrast curves by two orders of magnitude over the standard NRM configuration. This yields the ability to detect companions at $1\,\lambda/D$ separations to $10^{-4}$ contrast in approximately 10 seconds of integration time, or $5\times 10^{-6}$ in approximately 1 hour of integration time.

\begin{figure}[t] 
\centering
\includegraphics[width = 0.83\textwidth]{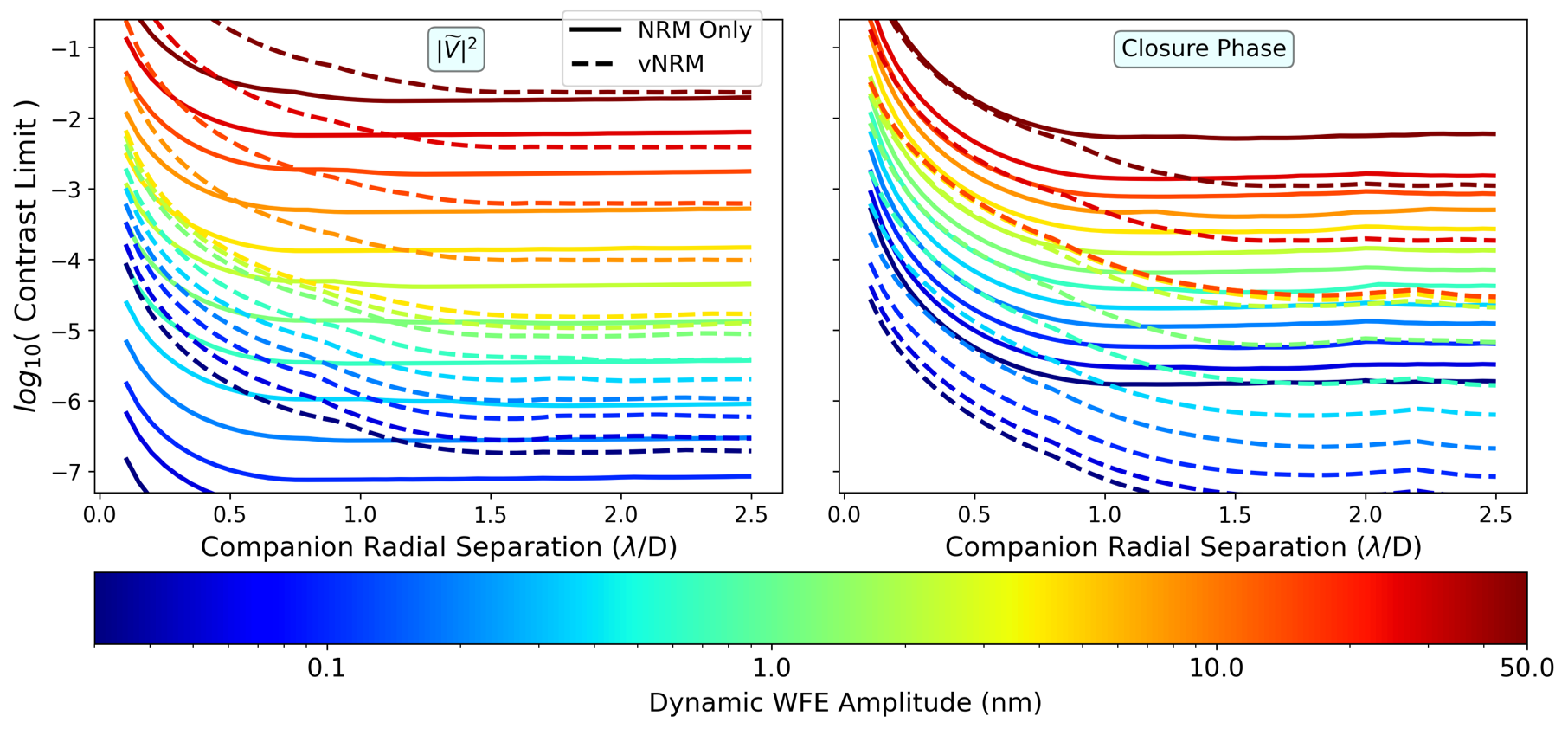}
\caption{Equivalent figure to Fig. \ref{fig:WFE_CC_static} but showing the sensitivity of the high-performance coronagraph to low-order dynamic wavefront error. Shows the WFE-limited $SNR=5$ contrast curves for squared visibilities (left) and closure phases (right).} \label{fig:LOWFE_CC_c2_No2nd} 
\end{figure}

We also repeat the same tests as in Sec. \ref{sec:WFE_general} to create the WFE-limited contrast curve for the high performance coronagraph. In this case, it turns out the visibility signals also has a dependence on the WFE amplitude. 
As the visibility signal of the vNRM to an off-axis companion is dependent on the contrast gain supplied by the coronagraph, $\sigma$\,radians of WFE allows roughly $1-\sigma^2$ of the on-axis flux to leak through the coronagraph, which then influences the contrast limit for the detectable companions by suppressing the companion signal. 
When incorporating this effect into the contrast curve, we arrive at Fig. \ref{fig:LOWFE_CC_c2_No2nd}, where the contrast curves for the vNRM are now comparable to, or reach deeper contrasts than, the standard NRM configuration. Figure \ref{fig:LOWFE_CC_c2_No2nd} shows the WFE-limited contrast curve in response to low-order dynamic WFE. With a high-performance coronagraph, the sensitivity to dynamic and static WFE is similar, as the performance in this case is limited by leakage through the coronagraph. The contrast curves in response to high-order WFE is approximately 1.5 orders of magnitude deeper than the low-order WFE case. These findings can be used to inform design and stability requirements to achieve a desired contrast performance with this technique.

\subsection{Future Directions} \label{sec:Future_work_discuss}
The scope of this paper is to introduce the vNRM system and investigate a selection of its fundamental properties in controlled scenarios. This 
is not sufficient to fully build expectations of the performance of an instrument based on this concept. 
In this section, we identify and discuss necessary future directions of inquiry to further develop the conceptual design, and we give initial comments of our speculation toward each topic. 

The most necessary aspect for the further development of hte vNRM 
will be optimizing the design of the non-redundant mask. 
As has been mentioned prior, many NRM designs have been used in other instruments, all of which utilize more subapertures to measure more baselines and closure triangles at more locations in the UV plane -- many of which yield a higher total throughput than our simulation \cite{2019AJ....157..249G, 2023PASP..135a5003S, 2023SPIE12680E..24L}. These mask designs can serve as initial configurations when optimizing mask parameters, prior to accounting for any obstructions in the entrance pupil, such as a secondary mirror, spiders, segments, etc. 
With the measurement of complex visibilities from additional baselines, additional high-order visibility observables, such as the biphase or closure amplitudes, can be used to further constrain the astrophysical scene. 
It is also worth mentioning that the light rejected by the NRM does not necessarily need to be wasted. By making the NRM a reflective/transmissive surface, the rejected light could be directed to another detector for potentially centering the on-axis light on the coronagraph, wavefront sensing, or some other observational technique, such as kernel phase interferometry. Alternatively, the use of jewel optics in the Lyot plane could dramatically improve the total throughput of the NRM aperture while retaining non-redundant baselines \cite{2025JATIS..11a5004T}. 

Lastly, with regards to the optimization of the NRM, it may be worth investigating the maximum allowable subaperture size. In this study, we set the subapertures size to be a factor of $2\pi$ smaller than the smallest difference in baseline vectors, preventing any cross-talk between baselines following the propagation into the UV plane. However, subaperture size is directly related to the NRM total throughput, so there exists a tradeoff between improvement to photon noise and uncertainty created by cross-talk between baselines. 
The size of the subapertures may also affect the visibilities' sensitivity to WFE, especially for higher-charge vortices with a secondary obstruction. 
In those scenarios, the complex amlplitude at the Lyot plane contains null regions where the field amplitude goes to zero with a phase flip, so subapertures that span the null become very sensitive to WFE. With smaller subapertures, it becomes less likely that a subaperture will cross a null, so the NRM will become more resilient against WFE. 

Following further development of the mask, the next necessary development pathway 
is the broadband behavior of the vNRM system. We mentioned previously that the contrast provided by a classical scalar vortex is significantly chromatic. However, there have been experiments showing that by adding a Roddier phase dimple, the scalar vortex becomes more achromatic\cite{2026arXiv260322537D}. It may also be possible for metasurface and PIAACMC coronagraphs to achieve broadband, low-IWA on-axis starlight suppression\cite{2023SPIE12680E..0QK, 2024OExpr..3247057P, 2026A&A...708A.144T}. 
However, even when assuming achromatic on-axis starlight suppression from the coronagraph, 
we must still contend with the the fact that the fringes generated by the NRM broaden with wavelength, decreasing the sensitivity to astrophysical signal at large spectral bandwidths. This can be counteracted by sampling the NRM interferogram with an integral field unit (IFU) to measure the spectrum at every point in the interferogram. The GPI NRM subsystem makes use of this architecture, and the addition of measurements at different wavelengths allows for an increased number of measurements and improved sensitivity\cite{2019AJ....157..249G}. 
As was mentioned in Sec. \ref{sec:sim_setup}, we simulated the photon flux of an NRM with an effective spectral bandwidth of $0.2\%$, corresponding to a spectral resolution of $R\sim500$. From this, we can be confident that the photon limit of the vNRM will remain relatively unchanged even when paired with an IFU with moderate spectral resolution. However, further investigation will be necessary to estimate any second-order effects of sampling the post-coronagraphic interferogram and how being fed into a spectrograph may affect the contrast limits. 

In addition to improving the veracity of the physical design of the vNRM system, it will also be necessary to develop expectations of the vNRM behavior for more realistic astrophysical sources. Beyond high-contrast companions, extended sources (e.g. proto-planetary disks, debris disks, etc.) are also of astrophysical interest at small angular scales. 
Not only will targets with different astrophysical morphologies generate dissimilar visibility signals to those of a point-source companion, but the stellar leakage of an on-axis target with non-zero spatial extent may ultimately affect the achievable photon limit attainable by the vNRM. 
There exist code packages that have been used to simulate the observations, including interferometric visibilities, of 
extended sources (e.g. MCFOST\cite{2006A&A...459..797P}, PMOIRED\cite{2022ascl.soft05001M}, etc.). 
It may be possible to make modifications to existing software to simulate observations of extended sources with a vNRM system as well. 

Additionally, as the vNRM samples visibilities in the Lyot plane, as was mentioned before, it will be necessary to investigate methods for data processing and image reconstruction for vNRM data. There exist various data reduction schemes for standard NRM data\cite{1999PhDT........19M, 2017ApJS..233....9S, 2019AJ....157..249G}. Our initial speculation is these data reduction pipelines will reproduce data from the post-coronagraph PSF and that accounting for the coronagraph transfer function will reproduce the astrophysical source, but this development is far beyond the scope of this stage of the project. 
Knowing that static WFE creates deviations from the reference visibility observables to create measurement error, it may also be possible to incorporate measured static wavefront error at the time of the exposure to compensate for these static deviations. This is similar in principle to PSF reconstruction efforts\cite{1997JOSAA..14.3057V, 2006A&A...457..359G}. 
Most data-reduction pipelines also include controlling for detector effects and noise. These were not simulated in this paper but could potentially affect the contrast performance of the vNRM. 
In this regard, another potential opportunity for the vNRM system arises when considering the high dynamic range of the astrophysical field during high-contrast observations. 
It has been observed that non-linear "brighter-fatter" detector effect can limit 
the contrast performance of the NRM on the JWST-IRIS instrument\cite{2024ApJ...963L...2S, 2025arXiv251009806D}. With a preceding coronagraph, the dynamic range of the intensity at the detector plane is decreased due to the on-axis light being suppressed. The vNRM would thus be less affected by this specific detector effect compared to a standard NRM architecture. 
Ultimately, it will never be possible in simulation to fully account for every physical discrepancy that may affect instrument performance prior to the creation of the instrument 
(e.g. segmented apertures, hardware imperfections and misalignments, pupil rotation and nutation, etc). Experimental validation of our simulated findings and conclusions remains as a key next step for advancing the TRL of the vNRM system. 

\section{Key Takeaways} \label{sec:conclusion}
Non-redundant aperture masking converts a full-aperture telescope into an interferometric array capable of detecting astrophysical features within the telescope's typical diffraction limit. However, this technique is typically limited to detecting features at a contrast of approximately $\Delta\text{mag}\approx 7$. By combining with an upstream vortex coronagraph, which is capable of effectively improving the contrast between on-axis and off-axis emission sources, into a vNRM system, we become capable of detecting significantly fainter companion sources. 

By comparing the performance of the vNRM system to a standard NRM configuration limited by photon noise, we showed a contrast improvement of $\sim 0.5$ orders of magnitude by including a vortex coronagraph in the case of a telescope with a secondary obstruction, or an improvement of 2.5 orders of magnitude for an off-axis telescope without a secondary obstruction. In the case that the observations with the system are WFE limited, the vNRM experiences greater levels of measurement error than the standard NRM, indicating the need for more stringent wavefront stability requirements for a vNRM system, as compared to those for a standard NRM system. 

\appendix    

\section{Complex Amplitudes in the Lyot Plane}
\label{sec:math_appendix}
In this appendix, we more explicitly derive the complex amplitude at the location of the aperture mask in the vNRM system. We begin with the vortex mask applied to the focal plane $\widetilde{E}_1$, and describe the propagation to the subsequent pupil plane: 
$$\widetilde{E}_{1, \text{post}}(\rho, \phi) = \iint \frac{R_\text{P}J_1(R_\text{P}\rho)}{\rho} \text{ exp}[im\phi] \text{ exp}[-i\rho r\text{cos}(\phi-\theta)]\rho d\rho d\phi$$
$$=R_\text{P}\int J_1(R_\text{P}\rho) \int \text{ exp}[im\phi] \text{ exp}[-i\rho r\text{cos}(\phi-\theta)]d\phi d\rho$$
The angular integral is now a form of the Bessel integral\cite{1998JOSAB..15.2226S} and is evaluated as: 
$$=2\pi R_\text{P} \text{ exp}[im\theta]\int J_1(R_\text{P}\rho) J_m(r\rho) d\rho.$$
This radial integral is now the form of a Weber-Schafheitlin integral\cite{2008MNRAS.384..515J}, which has the closed form solution\cite{2007tisp.book.....G}:
\begin{equation}
    =2\pi R_\text{P} \text{ exp}[im\theta]
    \cdot 
    \left\{
        \begin{array}{lr}
            0, & r < R_\text{P}\\
            \frac{R_\text{P}(m/2)!}{r^2((m/2) -1)!}F(\frac{m}{2}+1, 1-\frac{m}{2}; 2; \left(\frac{R_\text{P}}{r}\right)^2), &  r \geq R_\text{P}
        \end{array}
    \right\},
    \label{eqn:solve_WeberSchaf}
\end{equation}
where $F$ is the Gaussian hypergeometric function. The absence of amplitude within the unobscured pupil is the classical motivation for vortex coronagraphy\cite{2005OptL...30.3308F, 2005ApJ...633.1191M}. As was indicated in Ref. \citenum{2008MNRAS.384..515J}, diffraction from the secondary mirror causes the complex amplitude in the Lyot plane to be the sum of Eq. \ref{eqn:solve_WeberSchaf} for $R_p$ standing for the radius of the primary mirror and also for the radius of the secondary mirror, injecting light into the Lyot region. 
The complete complex amplitude in the Lyot plane is then given in Eq. \ref{eqn:lyot_full}.

For $m=$ 2, the entire Weber-Schafheitlin integral simplifies to a factor of $1/r^2$. For the charges of $m=$ 4, 6, and 8, we expand the hyper geometric function as:
\begin{equation}
    \begin{array}{lr}
        m=4, & F\left(3, -1; 2;\left(\frac{R_\text{P}}{r}\right)^2\right) = 1 - \frac{3}{2} \left(\frac{R_\text{P}}{r}\right)^2 \\
        m=6, & F\left(4, -2; 2;\left(\frac{R_\text{P}}{r}\right)^2\right) = 1 - 4 \left(\frac{R_\text{P}}{r}\right)^2 + \frac{10}{3} \left(\frac{R_\text{P}}{r}\right)^4\\
        m=8, & F\left(5, -3; 2;\left(\frac{R_\text{P}}{r}\right)^2\right) = 1 - \frac{15}{2} \left(\frac{R_\text{P}}{r}\right)^2 + 15 \left(\frac{R_\text{P}}{r}\right)^4 - \frac{35}{4} \left(\frac{R_\text{P}}{r}\right)^6.\\
    \end{array}
    \label{eqn:expand_Ffunc}
\end{equation}
The complex amplitude of the electric field inside of the Lyot region is described by these polynomials, multiplied by another factor of $1/r^2$ and some scaling constants. These show not only that the electric field incident to the NRM is non-constant, but also that the Lyot plane contains null regions for vortices with optical charge $m>2$. We do not derive it analytically, but numerical simulations show that these null regions change both shape and position in response to the incident light source moving off-axis, which has implications for the sensitivity of the vNRM subapertures to WFE and also to the signal of off-axis companions. 

\acknowledgments 
This work was supported by the Heising-Simons Foundation through grant \#2020-1821.

\bibliography{report} 

\begin{thebibliography}{10}

\bibitem{2021pdaa.book.....N}
{National Academies of Sciences} and Medicine, E.,  [{\em {Pathways to Discovery in Astronomy and Astrophysics for the 2020s}}{\nolinebreak\hspace{0.1em}]} (2021).

\bibitem{2021Msngr.182...38K}
{Kasper}, M., {Cerpa Urra}, N., {Pathak}, P., {Bonse}, M., {Nousiainen}, J., {Engler}, B., {Heritier}, C.~T., {Kammerer}, J., {Leveratto}, S., {Rajani}, C., {Bristow}, P., {Le Louarn}, M., {Madec}, P.-Y., {Str{\"o}bele}, S., {Verinaud}, C., {Glauser}, A., {Quanz}, S.~P., {Helin}, T., {Keller}, C., {Snik}, F., {Boccaletti}, A., {Chauvin}, G., {Mouillet}, D., {Kulcs{\'a}r}, C., and {Raynaud}, H.-F., ``{PCS {\textemdash} A Roadmap for Exoearth Imaging with the ELT},'' {\em The Messenger}~{\bf 182},  38--43 (Mar. 2021).

\bibitem{2005ApJ...633.1191M}
{Mawet}, D., {Riaud}, P., {Absil}, O., and {Surdej}, J., ``{Annular Groove Phase Mask Coronagraph},'' {\em \apj}~{\bf 633},  1191--1200 (Nov. 2005).

\bibitem{2005OptL...30.3308F}
{Foo}, G., {Palacios}, D.~M., and {Swartzlander}, Jr., G.~A., ``{Optical vortex coronagraph},'' {\em Optics Letters}~{\bf 30},  3308--3310 (Dec. 2005).

\bibitem{2008MNRAS.384..515J}
{Jenkins}, C., ``{Optical vortex coronagraphs on ground-based telescopes},'' {\em \mnras}~{\bf 384},  515--524 (Feb. 2008).

\bibitem{2009OExpr..17.1902M}
{Mawet}, D., {Serabyn}, E., {Liewer}, K., {Hanot}, C., {McEldowney}, S., {Shemo}, D., and {O'Brien}, N., ``{Optical Vectorial Vortex Coronagraphs using Liquid Crystal Polymers: theory, manufacturing and laboratory demonstration},'' {\em Optics Express}~{\bf 17},  1902--1918 (Feb. 2009).

\bibitem{2014ApJ...780...25E}
{Esposito}, T.~M., {Fitzgerald}, M.~P., {Graham}, J.~R., and {Kalas}, P., ``{Modeling Self-subtraction in Angular Differential Imaging: Application to the HD 32297 Debris Disk},'' {\em \apj}~{\bf 780},  25 (Jan. 2014).

\bibitem{2019A&A...631A.106S}
{Singh}, G., {Galicher}, R., {Baudoz}, P., {Dupuis}, O., {Ortiz}, M., {Potier}, A., {Thijs}, S., and {Huby}, E., ``{Active minimization of non-common path aberrations in long-exposure imaging of exoplanetary systems},'' {\em \aap}~{\bf 631},  A106 (Nov. 2019).

\bibitem{2022A&A...665A.136P}
{Potier}, A., {Mazoyer}, J., {Wahhaj}, Z., {Baudoz}, P., {Chauvin}, G., {Galicher}, R., and {Ruane}, G., ``{Increasing the raw contrast of VLT/SPHERE with the dark hole technique. II. On-sky wavefront correction and coherent differential imaging},'' {\em \aap}~{\bf 665},  A136 (Sept. 2022).

\bibitem{2024AJ....168...78W}
{Wallack}, N.~L., {Ruffio}, J.-B., {Ruane}, G., {Ren}, B.~B., {Xuan}, J.~W., {Villenave}, M., {Mawet}, D., {Stapelfeldt}, K., {Wang}, J.~J., {Liu}, M.~C., {Absil}, O., {Alvarez}, C., {Bae}, J., {Bond}, C., {Bottom}, M., {Calvin}, B., {Choquet}, {\'E}., {Christiaens}, V., {Cook}, T., {Femen{\'\i}a Castell{\'a}}, B., {Gomez Gonzalez}, C., {Guidi}, G., {Huby}, E., {Kastner}, J., {Knutson}, H.~A., {Meshkat}, T., {Ngo}, H., {Ragland}, S., {Reggiani}, M., {Ricci}, L., {Serabyn}, E., {Uyama}, T., {Williams}, J.~P., {Wizinowich}, P., {Zawol}, Z., {Zhang}, S., and {Zhu}, Z., ``{A Survey of Protoplanetary Disks Using the Keck/NIRC2 Vortex Coronagraph},'' {\em \aj}~{\bf 168},  78 (Aug. 2024).

\bibitem{2007SPIE.6691E..0AG}
{Give'on}, A., {Kern}, B., {Shaklan}, S., {Moody}, D.~C., and {Pueyo}, L., ``{Broadband wavefront correction algorithm for high-contrast imaging systems},'' in [{\em Astronomical Adaptive Optics Systems and Applications III}{\nolinebreak\hspace{0.1em}]},  {Tyson}, R.~K. and {Lloyd-Hart}, M., eds., {\em Society of Photo-Optical Instrumentation Engineers (SPIE) Conference Series} {\bf 6691},  66910A (Sept. 2007).

\bibitem{2017JATIS...3d9002M}
{Miller}, K., {Guyon}, O., and {Males}, J., ``{Spatial linear dark field control: stabilizing deep contrast for exoplanet imaging using bright speckles},'' {\em Journal of Astronomical Telescopes, Instruments, and Systems}~{\bf 3},  049002 (Oct. 2017).

\bibitem{2006dies.conf..553B}
{Baudoz}, P., {Boccaletti}, A., {Baudrand}, J., and {Rouan}, D., ``{The Self-Coherent Camera: a new tool for planet detection},'' in [{\em IAU Colloquium 200: Direct Imaging of Exoplanets: Science \& Techniques}{\nolinebreak\hspace{0.1em}]},  {Aime}, C. and {Vakili}, F., eds.,  553--558 (Jan. 2006).

\bibitem{2020A&A...635A.192P}
{Potier}, A., {Baudoz}, P., {Galicher}, R., {Singh}, G., and {Boccaletti}, A., ``{Comparing focal plane wavefront control techniques: Numerical simulations and laboratory experiments},'' {\em \aap}~{\bf 635},  A192 (Mar. 2020).

\bibitem{2019SPIE11117E..1FR}
{Ruane}, G., {Mawet}, D., {Riggs}, A.~J.~E., and {Serabyn}, E., ``{Scalar vortex coronagraph mask design and predicted performance},'' in [{\em Society of Photo-Optical Instrumentation Engineers (SPIE) Conference Series}{\nolinebreak\hspace{0.1em}]},  {\em Society of Photo-Optical Instrumentation Engineers (SPIE) Conference Series} {\bf 11117},  111171F (Sept. 2019).

\bibitem{2022SPIE12180E..25B}
{Belikov}, R., {Sirbu}, D., {Marx}, D., {Mejia Prada}, C., {Bendek}, E., {Pluzhnik}, E., {Bryson}, S., {Kern}, B., {Guyon}, O., {Fogarty}, K., {Knight}, J., {Wilson}, D., and {Hagopian}, J., ``{Laboratory demonstration of high contrast with the PIAACMC coronagraph on an obstructed and segmented aperture},'' in [{\em Space Telescopes and Instrumentation 2022: Optical, Infrared, and Millimeter Wave}{\nolinebreak\hspace{0.1em}]},  {Coyle}, L.~E., {Matsuura}, S., and {Perrin}, M.~D., eds., {\em Society of Photo-Optical Instrumentation Engineers (SPIE) Conference Series} {\bf 12180},  1218025 (Aug. 2022).

\bibitem{2026arXiv260322551D}
{Desai}, N., {Ruane}, G., {Redmond}, S., {Mawet}, D., {Serabyn}, E., and {Mennesson}, B., ``{Dimpled scalar vortex coronagraph laboratory demonstration},'' {\em arXiv e-prints} ,  arXiv:2603.22551 (Mar. 2026).

\bibitem{1958MNRAS.118..276J}
{Jennison}, R.~C., ``{A phase sensitive interferometer technique for the measurement of the Fourier transforms of spatial brightness distributions of small angular extent},'' {\em \mnras}~{\bf 118},  276 (Jan. 1958).

\bibitem{1986Natur.320..595B}
{Baldwin}, J.~E., {Haniff}, C.~A., {Mackay}, C.~D., and {Warner}, P.~J., ``{Closure phase in high-resolution optical imaging},'' {\em \nat}~{\bf 320},  595--597 (Apr. 1986).

\bibitem{2000PASP..112..555T}
{Tuthill}, P.~G., {Monnier}, J.~D., {Danchi}, W.~C., {Wishnow}, E.~H., and {Haniff}, C.~A., ``{Michelson Interferometry with the Keck I Telescope},'' {\em \pasp}~{\bf 112},  555--565 (Apr. 2000).

\bibitem{2011A&A...532A..72L}
{Lacour}, S., {Tuthill}, P., {Amico}, P., {Ireland}, M., {Ehrenreich}, D., {Huelamo}, N., and {Lagrange}, A.-M., ``{Sparse aperture masking at the VLT. I. Faint companion detection limits for the two debris disk stars HD 92945 and HD 141569},'' {\em \aap}~{\bf 532},  A72 (Aug. 2011).

\bibitem{1999PhDT........19M}
{Monnier}, J.~D., {\em {Infrared interferometry and spectroscopy of circumstellar envelopes}}, PhD thesis, University of California, Berkeley (Jan. 1999).

\bibitem{2006ApJ...650L.131L}
{Lloyd}, J.~P., {Martinache}, F., {Ireland}, M.~J., {Monnier}, J.~D., {Pravdo}, S.~H., {Shaklan}, S.~B., and {Tuthill}, P.~G., ``{Direct Detection of the Brown Dwarf GJ 802B with Adaptive Optics Masking Interferometry},'' {\em \apjl}~{\bf 650},  L131--L134 (Oct. 2006).

\bibitem{2008ApJ...678L..59I}
{Ireland}, M.~J. and {Kraus}, A.~L., ``{The Disk Around CoKu Tauri/4: Circumbinary, Not Transitional},'' {\em \apjl}~{\bf 678},  L59 (May 2008).

\bibitem{2010A&ARv..18..317A}
{Absil}, O. and {Mawet}, D., ``{Formation and evolution of planetary systems: the impact of high-angular resolution optical techniques},'' {\em \aapr}~{\bf 18},  317--382 (July 2010).

\bibitem{2019AJ....157..249G}
{Greenbaum}, A.~Z., {Cheetham}, A., {Sivaramakrishnan}, A., {Rantakyr{\"o}}, F.~T., {Duch{\^e}ne}, G., {Tuthill}, P., {De Rosa}, R.~J., {Oppenheimer}, R., {Macintosh}, B., {Ammons}, S.~M., {Bailey}, V.~P., {Barman}, T., {Bulger}, J., {Cardwell}, A., {Chilcote}, J., {Cotten}, T., {Doyon}, R., {Fitzgerald}, M.~P., {Follette}, K.~B., {Gerard}, B.~L., {Goodsell}, S.~J., {Graham}, J.~R., {Hibon}, P., {Hung}, L.-W., {Ingraham}, P., {Kalas}, P., {Konopacky}, Q., {Larkin}, J.~E., {Maire}, J., {Marchis}, F., {Marley}, M.~S., {Marois}, C., {Metchev}, S., {Millar-Blanchaer}, M.~A., {Morzinski}, K.~M., {Nielsen}, E.~L., {Palmer}, D., {Patience}, J., {Perrin}, M., {Poyneer}, L., {Pueyo}, L., {Rajan}, A., {Rameau}, J., {Sadakuni}, N., {Savransky}, D., {Schneider}, A.~C., {Song}, I., {Soummer}, R., {Thomas}, S., {Wallace}, J.~K., {Wang}, J.~J., {Ward-Duong}, K., {Wiktorowicz}, S., and {Wolff}, S., ``{Performance of the Gemini Planet Imager Non-redundant Mask and Spectroscopy of Two Close-separation Binaries: HR 2690 and HD
  142527},'' {\em \aj}~{\bf 157},  249 (June 2019).

\bibitem{2019JATIS...5a8001S}
{Sallum}, S. and {Skemer}, A., ``{Comparing nonredundant masking and filled-aperture kernel phase for exoplanet detection and characterization},'' {\em Journal of Astronomical Telescopes, Instruments, and Systems}~{\bf 5},  018001 (Jan. 2019).

\bibitem{2010JOSAA..27A.157S}
{Sauvage}, J.-F., {Mugnier}, L.~M., {Rousset}, G., and {Fusco}, T., ``{Analytical expression of long-exposure adaptive-optics-corrected coronagraphic image First application to exoplanet detection},'' {\em Journal of the Optical Society of America A}~{\bf 27},  A157 (Nov. 2010).

\bibitem{1992A&A...253..641T}
{Tallon}, M. and {Tallon-Bosc}, I., ``{The object-image relationship in Michelson stellar interferometry},'' {\em \aap}~{\bf 253},  641--645 (Jan. 1992).

\bibitem{por2018hcipy}
Por, E.~H., Haffert, S.~Y., Radhakrishnan, V.~M., Doelman, D.~S., Van~Kooten, M., and Bos, S.~P., ``{High Contrast Imaging for Python (HCIPy): an open-source adaptive optics and coronagraph simulator},'' in [{\em Adaptive Optics Systems VI}{\nolinebreak\hspace{0.1em}]},  {\em Proc. {{SPIE}}} {\bf 10703} (2018).

\bibitem{2010SPIE.7739E..14M}
{Mawet}, D., {Pueyo}, L., {Moody}, D., {Krist}, J., and {Serabyn}, E., ``{The Vector Vortex Coronagraph: sensitivity to central obscuration, low-order aberrations, chromaticism, and polarization},'' in [{\em Modern Technologies in Space- and Ground-based Telescopes and Instrumentation}{\nolinebreak\hspace{0.1em}]},  {Atad-Ettedgui}, E. and {Lemke}, D., eds., {\em Society of Photo-Optical Instrumentation Engineers (SPIE) Conference Series} {\bf 7739},  773914 (July 2010).

\bibitem{2024JATIS..10a5001D}
{Desai}, N., {Mawet}, D., {Serabyn}, E., {Ruane}, G., {Bertrou-Cantou}, A., {Llop-Sayson}, J., and {Eldorado Riggs}, A.~J., ``{Benefits of adding radial phase dimples on scalar coronagraph phase masks},'' {\em Journal of Astronomical Telescopes, Instruments, and Systems}~{\bf 10},  015001 (Jan. 2024).

\bibitem{2026arXiv260322537D}
{Desai}, N., {Ruane}, G., {Shanks}, D., {K{\"o}nig}, L., {Redmond}, S., and {Mennesson}, B., ``{Model validation and tolerancing of scalar vortex masks in the High Contrast Imaging Testbed (HCIT) facility},'' {\em arXiv e-prints} ,  arXiv:2603.22537 (Mar. 2026).

\bibitem{2018SPIE10698E..2VR}
{Riggs}, A.~J.~E., {Ruane}, G., {Sidick}, E., {Coker}, C., {Kern}, B.~D., and {Shaklan}, S.~B., ``{Fast linearized coronagraph optimizer (FALCO) I: a software toolbox for rapid coronagraphic design and wavefront correction},'' in [{\em Space Telescopes and Instrumentation 2018: Optical, Infrared, and Millimeter Wave}{\nolinebreak\hspace{0.1em}]},  {Lystrup}, M., {MacEwen}, H.~A., {Fazio}, G.~G., {Batalha}, N., {Siegler}, N., and {Tong}, E.~C., eds., {\em Society of Photo-Optical Instrumentation Engineers (SPIE) Conference Series} {\bf 10698},  106982V (Aug. 2018).

\bibitem{2023PASP..135a5003S}
{Sivaramakrishnan}, A., {Tuthill}, P., {Lloyd}, J.~P., {Greenbaum}, A.~Z., {Thatte}, D., {Cooper}, R.~A., {Vandal}, T., {Kammerer}, J., {Sanchez-Bermudez}, J., {Pope}, B. J.~S., {Blakely}, D., {Albert}, L., {Cook}, N.~J., {Johnstone}, D., {Martel}, A.~R., {Volk}, K., {Soulain}, A., {Artigau}, {\'E}., {Lafreni{\`e}re}, D., {Willott}, C.~J., {Parmentier}, S., {Ford}, K.~E.~S., {McKernan}, B., {Vila}, M.~B., {Rowlands}, N., {Doyon}, R., {Beaulieu}, M., {Desdoigts}, L., {Fullerton}, A.~W., {De Furio}, M., {Goudfrooij}, P., {Holfeltz}, S.~T., {LaMassa}, S., {Maszkiewicz}, M., {Meyer}, M.~R., {Perrin}, M.~D., {Pueyo}, L., {Sahlmann}, J., {Sohn}, S.~T., {Teixeira}, P.~S., and {Zheng}, S.-h., ``{The Near Infrared Imager and Slitless Spectrograph for the James Webb Space Telescope. IV. Aperture Masking Interferometry},'' {\em \pasp}~{\bf 135},  015003 (Jan. 2023).

\bibitem{2023SPIE12680E..24L}
{Lach}, M.~R., {Sallum}, S., {Banyal}, R., {Batalha}, N., {Blake}, G., {Brandt}, T., {Briesemeister}, Z., {Desai}, A., {Eisner}, J., {Fong}, W.-f., {Greene}, T., {Honda}, M., {Kain}, I., {Kilpatrick}, C., {de Kleer}, K., {Liu}, M., {Macintosh}, B., {Martinez}, R., {Mawet}, D., {Miles}, B., {Morley}, C., {de Pater}, I., {Powell}, D., {Sheehan}, P., {Skemer}, A., {Spilker}, J., {Stelter}, D., {Stone}, J., {Surya}, A., {Thirupathi}, S., {Wagner}, K., and {Zhou}, Y., ``{Recovering simulated planet and disk signals using SCALES aperture masking},'' in [{\em Society of Photo-Optical Instrumentation Engineers (SPIE) Conference Series}{\nolinebreak\hspace{0.1em}]},  {\em Society of Photo-Optical Instrumentation Engineers (SPIE) Conference Series} {\bf 12680},  1268024 (Oct. 2023).

\bibitem{2017ApJS..233....9S}
{Sallum}, S. and {Eisner}, J., ``{Data Reduction and Image Reconstruction Techniques for Non-redundant Masking},'' {\em \apjs}~{\bf 233},  9 (Nov. 2017).

\bibitem{2024AJ....167..139T}
{Tuchow}, N.~W., {Stark}, C.~C., and {Mamajek}, E., ``{HPIC: The Habitable Worlds Observatory Preliminary Input Catalog},'' {\em \aj}~{\bf 167},  139 (Mar. 2024).

\bibitem{2014ApJ...792...97M}
{Mawet}, D., {Milli}, J., {Wahhaj}, Z., {Pelat}, D., {Absil}, O., {Delacroix}, C., {Boccaletti}, A., {Kasper}, M., {Kenworthy}, M., {Marois}, C., {Mennesson}, B., and {Pueyo}, L., ``{Fundamental Limitations of High Contrast Imaging Set by Small Sample Statistics},'' {\em \apj}~{\bf 792},  97 (Sept. 2014).

\bibitem{2000JOSAA..17.1650S}
{Sch{\"o}ck}, M. and {Spillar}, E.~J., ``{Method for a quantitative investigation of the frozen flow hypothesis},'' {\em Journal of the Optical Society of America A}~{\bf 17},  1650--1658 (Sept. 2000).

\bibitem{2007MNRAS.378.1177B}
{Berdja}, A. and {Borgnino}, J., ``{Modelling the optical turbulence boiling and its effect on finite-exposure differential image motion},'' {\em \mnras}~{\bf 378},  1177--1186 (July 2007).

\bibitem{2014MNRAS.440.1925G}
{Guesalaga}, A., {Neichel}, B., {Cort{\'e}s}, A., {B{\'e}chet}, C., and {Guzm{\'a}n}, D., ``{Using the C$_{n}$$^{2}$ and wind profiler method with wide-field laser-guide-stars adaptive optics to quantify the frozen-flow decay},'' {\em \mnras}~{\bf 440},  1925--1933 (May 2014).

\bibitem{2022SPIE12185E..83C}
{Calvin}, B., {Fitzgerald}, M.~P., {van Kooten}, M. A.~M., {Fowler}, J., {Jensen-Clem}, R., {Gerard}, B.~L., and {Ragland}, S., ``{The use of spatial-temporal correlations to identify dynamic environmental changes affecting adaptive optics system performance},'' in [{\em Adaptive Optics Systems VIII}{\nolinebreak\hspace{0.1em}]},  {Schreiber}, L., {Schmidt}, D., and {Vernet}, E., eds., {\em Society of Photo-Optical Instrumentation Engineers (SPIE) Conference Series} {\bf 12185},  1218583 (Aug. 2022).

\bibitem{2013ApJ...776..130D}
{Dekany}, R., {Roberts}, J., {Burruss}, R., {Bouchez}, A., {Truong}, T., {Baranec}, C., {Guiwits}, S., {Hale}, D., {Angione}, J., {Trinh}, T., {Zolkower}, J., {Shelton}, J.~C., {Palmer}, D., {Henning}, J., {Croner}, E., {Troy}, M., {McKenna}, D., {Tesch}, J., {Hildebrandt}, S., and {Milburn}, J., ``{PALM-3000: Exoplanet Adaptive Optics for the 5 m Hale Telescope},'' {\em \apj}~{\bf 776},  130 (Oct. 2013).

\bibitem{2020SPIE11448E..0WM}
{Meeker}, S.~R., {Truong}, T.~N., {Roberts}, J.~E., {Shelton}, J.~C., {Fregoso}, S.~F., {Burruss}, R.~S., {Dekany}, R.~G., {Wallace}, J.~K., {Baker}, J.~W., {Heffner}, C.~M., {Mawet}, D., {Rykoski}, K.~M., {Tesch}, J.~A., and {Vasisht}, G., ``{Design and performance of the PALM-3000 3.5 kHz upgrade},'' in [{\em Adaptive Optics Systems VII}{\nolinebreak\hspace{0.1em}]},  {Schreiber}, L., {Schmidt}, D., and {Vernet}, E., eds., {\em Society of Photo-Optical Instrumentation Engineers (SPIE) Conference Series} {\bf 11448},  114480W (Dec. 2020).

\bibitem{2024arXiv240713019K}
{Kueny}, J.~K., {Van Gorkom}, K., {Kautz}, M., {Haffert}, S., {Males}, J.~R., {Hedglen}, A., {Close}, L., {McEwen}, E., {Li}, J., {Long}, J.~D., {Foster}, W., {Pearce}, L., {McLeod}, A., {Lumbres}, J., {Guyon}, O., and {Liberman}, J., ``{MagAO-X Phase II Upgrades: Implementation and First On-Sky Results of a New Post-AO 1000 Actuator Deformable Mirror},'' {\em arXiv e-prints} ,  arXiv:2407.13019 (July 2024).

\bibitem{2024arXiv240719188L}
{Lozi}, J., {Ahn}, K., {Blue}, H., {Chun}, A., {Clergeon}, C., {Deo}, V., {Guyon}, O., {Hattori}, T., {Minowa}, Y., {Nishiyama}, S., {Ono}, Y., {Oya}, S., {Takagi}, Y., {Vievard}, S., and {Vincent}, M., ``{AO3k at Subaru: First on-sky results of the facility extreme-AO},'' {\em arXiv e-prints} ,  arXiv:2407.19188 (July 2024).

\bibitem{2022SPIE12180E..0VK}
{Knight}, J.~S. and {Lightsey}, P.~A., ``{Webb Telescope imaging performance},'' in [{\em Space Telescopes and Instrumentation 2022: Optical, Infrared, and Millimeter Wave}{\nolinebreak\hspace{0.1em}]},  {Coyle}, L.~E., {Matsuura}, S., and {Perrin}, M.~D., eds., {\em Society of Photo-Optical Instrumentation Engineers (SPIE) Conference Series} {\bf 12180},  121800V (Aug. 2022).

\bibitem{2025SPIE13623E..0CE}
{Eegholm}, B.~H., {Johnson}, J., {Miller}, P., {Abel}, J., {Bolcar}, M., {Castle}, B., {Glebov}, B., {Jurling}, A., {McCarthy}, P., {Michaels}, E., {Paine}, S., {O'Sullivan}, M., {Gorski}, K., and {Smith}, J.~S., ``{Optical performance of Roman optical telescope assembly},'' in [{\em UV/Optical/IR Space Telescopes and Instruments: Innovative Technologies and Concepts XII}{\nolinebreak\hspace{0.1em}]},  {Arenberg}, J.~W. and {Stahl}, H.~P., eds., {\em Society of Photo-Optical Instrumentation Engineers (SPIE) Conference Series} {\bf 13623},  136230C (Sept. 2025).

\bibitem{2017SPIE10400E..0OE}
{Eldorado Riggs}, A.~J., {Zimmerman}, N.~T., {Nemati}, B., and {Krist}, J., ``{Shaped pupil coronagraph design improvements for the WFIRST coronagraph instrument},'' in [{\em Society of Photo-Optical Instrumentation Engineers (SPIE) Conference Series}{\nolinebreak\hspace{0.1em}]},  {Shaklan}, S., ed., {\em Society of Photo-Optical Instrumentation Engineers (SPIE) Conference Series} {\bf 10400},  104000O (Sept. 2017).

\bibitem{2018SPIE10703E..59L}
{Lozi}, J., {Guyon}, O., {Jovanovic}, N., {Goebel}, S., {Pathak}, P., {Skaf}, N., {Sahoo}, A., {Norris}, B., {Martinache}, F., {N'Diaye}, M., {Mazin}, B., {Walter}, A.~B., {Tuthill}, P., {Kudo}, T., {Kawahara}, H., {Kotani}, T., {Ireland}, M., {Cvetojevic}, N., {Huby}, E., {Lacour}, S., {Vievard}, S., {Groff}, T.~D., {Chilcote}, J.~K., {Kasdin}, J., {Knight}, J., {Snik}, F., {Doelman}, D., {Minowa}, Y., {Clergeon}, C., {Takato}, N., {Tamura}, M., {Currie}, T., {Takami}, H., and {Hayashi}, M., ``{SCExAO, an instrument with a dual purpose: perform cutting-edge science and develop new technologies},'' in [{\em Adaptive Optics Systems VI}{\nolinebreak\hspace{0.1em}]},  {Close}, L.~M., {Schreiber}, L., and {Schmidt}, D., eds., {\em Society of Photo-Optical Instrumentation Engineers (SPIE) Conference Series} {\bf 10703},  1070359 (July 2018).

\bibitem{2026A&A...708A.144T}
{Tonucci}, E., {Haffert}, S.~Y., {Foster}, W.~B., {Males}, J.~R., {Guyon}, O., {Close}, L.~M., {Van Gorkom}, K., {Hedglen}, A.~D., {Johnson}, P.~T., {Kautz}, M.~Y., {Kueny}, J.~K., {Li}, J., {Liberman}, J., {Long}, J.~D., {Lumbres}, J., {Mars}, M., {McEwen}, E.~A., {McLeod}, A., {Pearce}, L.~A., {Schatz}, L., and {Twitchell}, K., ``{Phase-Induced Amplitude Apodization Complex Mask Coronagraph (PIAACMC) on-sky demonstration with MagAO-X},'' {\em \aap}~{\bf 708},  A144 (Apr. 2026).

\bibitem{2011OptL...36.1506M}
{Mawet}, D., {Serabyn}, E., {Wallace}, J.~K., and {Pueyo}, L., ``{Improved high-contrast imaging with on-axis telescopes using a multistage vortex coronagraph},'' {\em Optics Letters}~{\bf 36},  1506 (Apr. 2011).

\bibitem{2016OptCo.379...64S}
{Serabyn}, E., {Liewer}, K., and {Mawet}, D., ``{Laboratory demonstration of a dual-stage vortex coronagraph},'' {\em Optics Communications}~{\bf 379},  64--67 (Nov. 2016).

\bibitem{2025JATIS..11a5004T}
{Taras}, A.~K., {Piroscia}, G., and {Tuthill}, P., ``{Jewel Optics I: non-redundant Fizeau beam combination without the guilt},'' {\em Journal of Astronomical Telescopes, Instruments, and Systems}~{\bf 11},  015004 (Jan. 2025).

\bibitem{2023SPIE12680E..0QK}
{K{\"o}nig}, L., {Palatnick}, S., {Desai}, N., {Absil}, O., {Millar-Blanchaer}, M., and {Mawet}, D., ``{Metasurface-based scalar vortex phase mask in pursuit of 1e-10 contrast},'' in [{\em Society of Photo-Optical Instrumentation Engineers (SPIE) Conference Series}{\nolinebreak\hspace{0.1em}]},  {\em Society of Photo-Optical Instrumentation Engineers (SPIE) Conference Series} {\bf 12680},  126800Q (Oct. 2023).

\bibitem{2024OExpr..3247057P}
{Palatnick}, S., {Millar-Blanchaer}, M.~A., {Wallace}, J.~K., {John}, D.~D., {Moore}, A., and {Wenger}, T., ``{Achromatizing photolithographically patterned metasurfaces with arbitrary, variable unit cell size},'' {\em Optics Express}~{\bf 32},  47057 (Dec. 2024).

\bibitem{2006A&A...459..797P}
{Pinte}, C., {M{\'e}nard}, F., {Duch{\^e}ne}, G., and {Bastien}, P., ``{Monte Carlo radiative transfer in protoplanetary disks},'' {\em \aap}~{\bf 459},  797--804 (Dec. 2006).

\bibitem{2022ascl.soft05001M}
{M{\'e}rand}, A., ``{PMOIRED: Parametric Modeling of Optical Interferometric Data}.'' Astrophysics Source Code Library, record ascl:2205.001 (May 2022).

\bibitem{1997JOSAA..14.3057V}
{Veran}, J.-P., {Rigaut}, F., {Maitre}, H., and {Rouan}, D., ``{Estimation of the adaptive optics long-exposure point-spread function using control loop data.},'' {\em Journal of the Optical Society of America A}~{\bf 14},  3057--3069 (Nov. 1997).

\bibitem{2006A&A...457..359G}
{Gendron}, E., {Cl{\'e}net}, Y., {Fusco}, T., and {Rousset}, G., ``{New algorithms for adaptive optics point-spread function reconstruction},'' {\em \aap}~{\bf 457},  359--363 (Oct. 2006).

\bibitem{2024ApJ...963L...2S}
{Sallum}, S., {Ray}, S., {Kammerer}, J., {Sivaramakrishnan}, A., {Cooper}, R., {Greebaum}, A.~Z., {Thatte}, D., {De Furio}, M., {Factor}, S.~M., {Meyer}, M.~R., {Stone}, J.~M., {Carter}, A., {Biller}, B., {Hinkley}, S., {Skemer}, A., {Su{\'a}rez}, G., {Leisenring}, J.~M., {Perrin}, M.~D., {Kraus}, A.~L., {Absil}, O., {Balmer}, W.~O., {Betti}, S.~K., {Boccaletti}, A., {Bonavita}, M., {Bonnefoy}, M., {Booth}, M., {Bowler}, B.~P., {Briesemeister}, Z.~W., {Bryan}, M.~L., {Calissendorff}, P., {Cantalloube}, F., {Chauvin}, G., {Chen}, C.~H., {Choquet}, E., {Christiaens}, V., {Cugno}, G., {Currie}, T., {Danielski}, C., {Dupuy}, T.~J., {Faherty}, J.~K., {Fitzgerald}, M.~P., {Fortney}, J.~J., {Franson}, K., {Girard}, J.~H., {Grady}, C.~A., {Gonzales}, E.~C., {Henning}, T., {Hines}, D.~C., {Hoch}, K. K.~W., {Hood}, C.~E., {Howe}, A.~R., {Janson}, M., {Kalas}, P., {Kennedy}, G.~M., {Kenworthy}, M.~A., {Kervella}, P., {Kitzmann}, D., {Kuzuhara}, M., {Lagrange}, A.-M., {Lagage}, P.-O., {Lawson}, K., {Lazzoni}, C., {Lew},
  B. W.~P., {Liu}, M.~C., {Liu}, P., {Llop-Sayson}, J., {Lloyd}, J.~P., {Lueber}, A., {Macintosh}, B., {Manjavacas}, E., {Marino}, S., {Marley}, M.~S., {Marois}, C., {Martinez}, R.~A., {Matthews}, B.~C., {Matthews}, E.~C., {Mawet}, D., {Mazoyer}, J., {McElwain}, M.~W., {Metchev}, S., {Miles}, B.~E., {Millar-Blanchaer}, M.~A., {Molliere}, P., {Moran}, S.~E., {Morley}, C.~V., {Mukherjee}, S., {Palma-Bifani}, P., {Pantin}, E., {Patapis}, P., {Petrus}, S., {Pueyo}, L., {Quanz}, S.~P., {Quirrenbach}, A., {Rebollido}, I., {Redai}, J.~A., {Ren}, B.~B., {Rickman}, E., {Samland}, M., {Sargent}, B.~A., {Schlieder}, J.~E., {Schneider}, G., {Stapelfeldt}, K.~R., {Sutlieff}, B.~J., {Tamura}, M., {Tan}, X., {Theissen}, C.~A., {Uyama}, T., {Vigan}, A., {Vasist}, M., {Vos}, J.~M., {Wagner}, K., {Wang}, J.~J., {Ward-Duong}, K., {Whiteford}, N., {Wolff}, S.~G., {Worthen}, K., {Wyatt}, M.~C., {Ygouf}, M., {Zhang}, X., {Zhang}, K., {Zhang}, Z., {Zhou}, Y., and {Zurlo}, A., ``{The JWST Early Release Science Program for Direct
  Observations of Exoplanetary Systems. IV. NIRISS Aperture Masking Interferometry Performance and Lessons Learned},'' {\em \apjl}~{\bf 963},  L2 (Mar. 2024).

\bibitem{2025arXiv251009806D}
{Desdoigts}, L., {Pope}, B., {Charles}, M., {Tuthill}, P., {Blakely}, D., {Johnstone}, D., {Ray}, S., {Sivaramakrishnan}, A., {Kammerer}, J., {Thatte}, D., and {Cooper}, R., ``{AMIGO: a Data-Driven Calibration of the JWST Interferometer},'' {\em arXiv e-prints} ,  arXiv:2510.09806 (Oct. 2025).

\bibitem{1998JOSAB..15.2226S}
{Sacks}, Z.~S., {Rozas}, D., and {Swartzlander}, Jr., G.~A., ``{Holographic formation of optical-vortex filaments},'' {\em Journal of the Optical Society of America B Optical Physics}~{\bf 15},  2226--2234 (Aug. 1998).

\bibitem{2007tisp.book.....G}
{Gradshteyn}, I.~S., {Ryzhik}, I.~M., {Jeffrey}, A., and {Zwillinger}, D.,  [{\em {Table of Integrals, Series, and Products}}{\nolinebreak\hspace{0.1em}]}, Elsevier Inc. (2007).

\end{thebibliography}
\bibliographystyle{spiebib} 

\end{document}